\documentstyle[12pt]{article}

\input epsf

\begin{document}

\newtheorem{theorem}{Theorem}[section]
\newtheorem{lemma}{Lemma}[section]
\newtheorem{corollary}{Corollary}[section]
\newtheorem{conjecture}{Conjecture}[section]
\newtheorem{example}{Example}[section]
\newtheorem{definition}{Definition}[section]
\newtheorem{postulate}{Postulate}[section]

\def\odt{{\textstyle{1\over 2}}}
\def\maxarg{\mathop{\rm maxarg}}          % maxarg
\def\minarg{\mathop{\rm minarg}}          % minarg

%\runninghead{Theories of Everything}

%\begin{titlepage}

\hfill Technical Report IDSIA-20-00, Version 2.0; 20 Dec 2000

%\hfill quant-ph/0011122
\hfill Minor revision of Version 1.0 \cite{Schmidhuber:00version1}, quant-ph/0011122

\vspace{2cm}
\begin{center}
{\LARGE ALGORITHMIC THEORIES OF EVERYTHING}

\end{center}
\vspace{0.3cm}

\begin{center}

J\"{u}rgen Schmidhuber

{\it IDSIA, Galleria 2, 6928 Manno (Lugano), Switzerland}

{\tt juergen@idsia.ch - http://www.idsia.ch/\~{ }juergen}

\end{center}

%\newpage

\begin{abstract}

The probability distribution $P$ from which the history of our universe
is sampled represents a theory of everything or TOE. We assume $P$ is
formally describable. Since most (uncountably many) distributions are not,
this imposes a strong inductive bias. We show that $P(x)$ is small for
any universe $x$ lacking a short description, and study the spectrum of
TOEs spanned by two $P$s, one reflecting the most compact constructive
descriptions, the other the fastest way of computing everything. The
former derives from generalizations of traditional computability,
Solomonoff's algorithmic probability, Kolmogorov complexity, and objects
more random than Chaitin's Omega, the latter from Levin's universal
search and a natural resource-oriented postulate: the cumulative prior
probability of all $x$ incomputable within time $t$ by this optimal
algorithm should be $1/t$.  Between both $P$s we find a universal
cumulatively enumerable measure that dominates traditional enumerable
measures; any such CEM must assign low probability to any universe lacking
a short enumerating program. We derive $P$-specific consequences for
evolving observers, inductive reasoning, quantum physics, philosophy,
and the expected duration of our universe.

\vspace{1.cm}
\hfill 10 theorems, 50 pages, 100 references, 20000 words 

\end{abstract}

\vspace{0.4cm}


\vspace{0.4cm}
\noindent
{\bf Note:} 
{\em 
This is a slightly revised version of a recent preprint
\cite{Schmidhuber:00version1}.  The essential results should be of
interest from a purely theoretical point of view independent of the
motivation through formally describable universes.  To get to the
meat of the paper, skip the introduction and go immediately to 
Subsection \ref{outline} which provides a condensed outline of the main 
theorems. 
}

%\end{titlepage}

\tableofcontents

\vspace{1cm}

\section{Introduction to Describable Universes}
\label{introduction}

An object $X$ is formally describable if a finite amount of information
completely describes $X$ and only $X$.  More to the point, $X$ should be
representable by a possibly infinite bitstring $x$ such that there is a
finite, possibly never halting program $p$ that computes $x$ and nothing
but $x$ in a way that modifies each output bit at most finitely many
times; that is, each finite beginning of $x$ eventually {\em converges}
and ceases to change. Definitions \ref{convergence}-\ref{describability}
will make this precise, and Sections \ref{preliminaries}-\ref{descriptive}
will clarify that this constructive notion of formal describability
is less restrictive than the traditional notion of computability
\cite{Turing:36}, mainly because we do not insist on the existence
of a halting program that computes an upper bound of the convergence
time of $p$'s $n$-th output bit. Formal describability thus pushes
constructivism \cite{Brouwer:07,Beeson:85} to the extreme, barely
avoiding the nonconstructivism embodied by even less restrictive
concepts of describability (compare computability {\em in the limit}
\cite{Gold:65,Putnam:65,Freyvald:74} and $\Delta^0_n$-describability
\cite{Rogers:67}\cite[p. 46-47]{LiVitanyi:97}).  The results in
Sections \ref{preliminaries}-\ref{coding} will exploit the additional
degrees of freedom gained over traditional computability, while Section
\ref{temporalcomp} will focus on another extreme, namely, the fastest 
way of computing all computable objects.

Among the formally describable things are the contents of all books
ever written, all proofs of all theorems, the infinite decimal
expansion of $\sqrt{17}$, and the enumerable ``number of wisdom''
$\Omega$ \cite{Chaitin:87,Slaman:99,Calude:00,Solovay:00}.  Most real
numbers, however, are not individually describable, because there are
only countably many finite descriptions, yet uncountably many reals,
as observed by Cantor in 1873 \cite{Cantor:1874}.  It is easy though to
write a never halting program that computes all finite prefixes of all
real numbers. In this sense certain sets seem describable while most of
their elements are not.

What about our universe, or more precisely, its entire past and future
history? Is it individually describable by a finite sequence of bits,
just like a movie stored on a compact disc, or a never ending evolution
of a virtual reality determined by a finite algorithm?  If so, then
it is very special in a certain sense, just like the comparatively few
describable reals are special.

\begin{example}[Pseudorandom universe]
\label{universe}
{\em
Let $x$ be an infinite sequence of finite bitstrings $x^1,x^2,\ldots$
representing the history of some discrete universe, where $x^k$ represents
the state of the universe at discrete time step $k$, and $x^1$ the
``Big Bang'' (compare \cite{Schmidhuber:97brauer}).  Suppose there
is a finite algorithm $A$ that computes $x^{k+1}$ ($k \geq 1$)  from
$x^{k}$ and additional information $noise^k$ (this may require numerous
computational steps of $A$, that is, ``local'' time of the universe may
run comparatively slowly).  Assume that $noise^k$ is not truly random
but calculated by invoking a finite pseudorandom generator subroutine \cite{Allender:89}.
Then $x$ is describable because it has a finite constructive description.
} 
\end{example}
Contrary to a widely spread misunderstanding, quantum physics, quantum
computation (e.g., \cite{Bennett:00,Deutsch:97,Penrose:89}) and Heisenberg's 
uncertainty principle do not rule out that our own universe's history is of
the type exemplified above.  It might be computable by a discrete
process approximated by Schr\"{o}dinger's continuous wave function,
where $noise^k$  determines the ``collapses'' of the wave function.
Since we prefer simple, formally describable explanations over complex,
nondescribable ones, we assume the history of our universe has a finite
description indeed.

This assumption has dramatic consequences.  For instance, because we
know that our future lies among the few (countably many) describable
futures, we can ignore uncountably many nondescribable ones.  Can we
also make more specific predictions?  Does it make sense to say some
describable futures are necessarily more likely than others?  To answer
such questions we will examine possible probability
distributions on possible futures, assuming that not only the histories
themselves but also their probabilities are formally describable. Since
most (uncountably many) real-valued probabilities are {\em not}, this
assumption --- against which there is no physical evidence --- actually
represents a major inductive bias,  which turns out to be strong enough to
explain certain hitherto unexplained aspects of our world.

\begin{example}[In which universe am I?]
\label{whereamI}
{\em
Let $h(y)$ represent a property of any possibly infinite
bitstring $y$, say, $h(y)=1$ if $y$ represents the history of a
universe inhabited by a particular observer 
(say, yourself)
and $h(y)=0$ otherwise.  According to the weak anthropic principle
\cite{Carter:74,BarrowTipler:86}, the conditional probability of finding 
yourself in a universe compatible with your existence equals 1.  
But there may be many $y$'s satisfying $h(y)=1$.
What is the probability that $y=x$, where $x$ is a particular
universe satisfying $h(x)=1$? According to Bayes,
\begin{equation}
P(x=y \mid h(y)=1) 
= \frac{P(h(y)=1 \mid x=y) P(x = y)} {\sum_{z:h(z)=1}  P(z)} 
\propto P(x) 
\label{bayes1}
\end{equation}
where $P(A \mid B)$ denotes the probability of $A$, given knowledge of $B$, 
and the denominator is just a normalizing constant.
So the probability of finding yourself in universe $x$
is essentially determined by $P(x)$, the
{\em prior probability} of $x$.  
}
\end{example}

Each prior $P$ stands for a particular ``theory of everything'' or TOE.
Once we know something about $P$ we can start making informed predictions.
Parts of this paper deal with the question: what are plausible properties of $P$?
One very plausible assumption is that $P$ is
{\em approximable} for all finite 
prefixes $\bar{x}$ of $x$ in the following sense.
There exists a possibly never halting computer which outputs a sequence of numbers
$T(t,\bar{x})$ at discrete times $t=1,2,\ldots$ in response to input
$\bar{x}$  such that for each real $\epsilon > 0$ there exists a finite
time $t_0$ such that for all $t \geq t_0$:
\begin{equation}
\mid P(\bar{x}) - T(t,\bar{x}) \mid < \epsilon .
\label{approx}
\end{equation}
Approximability in this sense is essentially equivalent to formal describability
(Lemma \ref{describabilityapproximability} will make this more precise).
We will show (Section \ref{coding})
that the mild assumption above adds enormous
predictive power to the weak anthropic principle: it makes universes
describable by short algorithms immensely more likely than others.
Any particular universe evolution is highly unlikely if it is determined
not only by simple physical laws but also by additional truly random or
noisy events. To a certain extent, this will justify ``Occam's razor''
(e.g., \cite{Blumer:87}) which expresses the ancient preference of
simple solutions over complex ones, and which is widely accepted not
only in physics and other inductive sciences, but even in the fine arts
\cite{Schmidhuber:97art}.

All of this will require an extension of earlier work
on Solomonoff's algorithmic probability, universal priors,
Kolmogorov complexity (or algorithmic information), and their refinements
\cite{Kolmogorov:65,Solomonoff:64,Chaitin:69,Zvonkin:70,Levin:73a,Levin:74,Gacs:74,Chaitin:75,Gacs:83,Schnorr:73,Solomonoff:78,Chaitin:87,Barzdin:88,Cover:89,Uspensky:92,LiVitanyi:97}.
We will prove several theorems concerning approximable and enumerable
objects and probabilities (Sections \ref{preliminaries}-\ref{coding};
see outline below). These theorems shed light on the structure of {\em
all} formally describable objects and extend traditional computability
theory; hence they should also be of interest without motivation through
describable universes.

The calculation of the subjects of these theorems, however, may
occasionally require excessive time, itself often not even computable
in the classic sense.  This will eventually motivate a shift of
focus on the temporal complexity of ``computing everything'' (Section
\ref{temporalcomp}).  If you were to sit down and write a program that
computes all possible universes, which would be the best way of doing so?
Somewhat surprisingly, a modification of Levin Search \cite{Levin:73}
can simultaneously compute all computable universes in an interleaving
fashion that outputs each individual universe as quickly as its fastest
algorithm running just by itself, save for a constant factor independent
of the universe's size. This suggests a more restricted TOE that singles
out those infinite universes computable with countable time and space
resources, and a natural resource-based prior measure $S$ on them.
Given this ``speed prior'' $S$, we will show that the most likely
continuation of a given observed history is computable by a fast and
short algorithm (Section \ref{Sbasedinference}).

The $S$-based TOE will provoke quite specific prophecies
concerning our own universe (Section \ref{Spredictions}).  For instance,
the probability that it will last $2^n$ times longer than it has lasted
so far is at most $2^{-n}$. Furthermore, all apparently random events,
such as beta decay or collapses of Schr\"{o}dinger's wave function of the
universe, actually must exhibit yet unknown, possibly nonlocal, regular
patterns reflecting subroutines (e.g., pseudorandom generators) of our
universe's algorithm that are not only short but also fast.

\subsection{Outline of Main Results}
\label{outline}

Some of the novel results herein may be of interest to
theoretical computer scientists and mathematicians (Sections
\ref{preliminaries}-\ref{temporalcomp}), some to researchers
in the fields of machine learning and inductive inference
(the science of making predictions based on observations,
e.g., \ref{temporalcomp}-\ref{consequences}), some to
physicists (e.g., \ref{temporalcomp}-\ref{conclusion}), some to
philosophers (e.g., \ref{consequences}-\ref{conclusion}).  Sections
\ref{consequences}-\ref{conclusion} might help those usually uninterested
in technical details to decide whether they would also like to delve
into the more formal Sections \ref{preliminaries}-\ref{temporalcomp}.
In what follows, we summarize the main contributions and 
provide pointers to the most important theorems.

Section \ref{preliminaries} introduces universal Turing Machines
(TMs) more general than those considered in previous related work:
unlike traditional TMs, {\em General} TMs or GTMs may edit their
previous outputs (compare inductive TMs \cite{Burgin:83}),
and {\em Enumerable Output Machines} (EOMs) may do
this provided the output does not decrease lexicographically.  We will
define: a formally describable object $x$ has a finite, never halting
GTM program that computes $x$ such that each output bit is revised at
most finitely many times; that is, each finite prefix of $x$ eventually
{\em stabilizes} (Defs. \ref{convergence}-\ref{describability});
describable {\em functions} can be implemented by such programs
(Def. \ref{describablefns}); weakly decidable {\em problems} have
solutions computable by never halting programs whose output is wrong for
at most finitely many steps (Def. \ref{weak}).  Theorem \ref{fluctuate}
generalizes the halting problem by demonstrating that it is not weakly
decidable whether a finite string is a description of a describable
object (compare a related result for analytic TMs by Hotz, Vierke and
Schieffer \cite{Hotz:95}).

Section \ref{descriptive} generalizes the traditional
concept of Kolmogorov complexity or algorithmic information
\cite{Kolmogorov:65,Solomonoff:64,Chaitin:69} of finite $x$ (the
length of the shortest halting program computing $x$) to the case
of objects describable by nonhalting programs on EOMs and GTMs
(Defs. \ref{generalizedK}-\ref{generalizedKm}).  It is shown that
the generalization for EOMs is describable, but the one for GTMs is
not (Theorem \ref{KGnot}).  Certain objects are much more compactly
encodable on EOMs than on traditional {\em monotone} TMs, and Theorem
\ref{z} shows that there are also objects with short GTM descriptions
yet incompressible on EOMs and therefore ``more random'' than Chaitin's
$\Omega$ \cite{Chaitin:87}, the halting probability of a TM
with random input, which is incompressible only on monotone TMs.
This yields a natural TM type-specific complexity hierarchy expressed
by Inequality (\ref{hierarchy}).

Section \ref{measures} discusses probability distributions on
describable objects as well as the nondescribable {\em convergence
probability} of a GTM (Def. \ref{upsilon}).  It also introduces
describable (semi)measures as well as {\em cumulatively enumerable}
measures (CEMs, Def. \ref{cem}), where the cumulative probability
of all strings lexicographically greater than a given string $x$
is EOM-computable or enumerable. Theorem \ref{universalCEM} shows
that there is a universal CEM that dominates all other CEMs, in the
sense that it assigns higher probability to any finite $y$, save for
a constant factor independent of $y$. This probability is shown to be
equivalent to the probability that an EOM whose input bits are chosen
randomly produces an output starting with $y$ (Corollary \ref{muE<mu0}
and Lemma \ref{mu0<muE}).  The nonenumerable universal CEM also dominates
enumerable priors studied in previous work by Solomonoff, Levin and others
\cite{Solomonoff:64,Zvonkin:70,Levin:74,Gacs:74,Chaitin:75,Gacs:83,Schnorr:73,Solomonoff:78,Chaitin:87,LiVitanyi:97}.
Theorem \ref{nouniversal} shows that there is no universal approximable
measure (proof by M. Hutter).

Section \ref{coding} establishes relationships between generalized
Kolmogorov complexity and generalized algorithmic probability, extending
previous work on enumerable semimeasures by Levin, G\'{a}cs, and others
\cite{Zvonkin:70,Levin:74,Gacs:74,Chaitin:75,Gacs:83,LiVitanyi:97}.
For instance, Theorem \ref{K-KP} shows that the universal CEM assigns
a probability to each enumerable object proportional to $\frac{1}{2}$
raised to the power of the length of its minimal EOM-based description,
times a small corrective factor. Similarly, objects with approximable
probabilities yet without very short descriptions on GTMs are necessarily
very unlikely {\em a priori} (Theorems \ref{KG-KPT} and \ref{KmG-Kmu}).
Additional suspected links between generalized Kolmogorov complexity and
probability are expressed in form of Conjectures \ref{KM?KP}-\ref{KG?KP}.

Section \ref{temporalcomp} addresses issues of temporal complexity
ignored in the previous sections on describable universe histories
(whose computation may require excessive time without recursive
bounds).  In Subsection \ref{FAST}, Levin's universal search algorithm
\cite{Levin:73,Levin:84} (which takes into account program runtime in
an optimal fashion) is modified to obtain the fastest way of computing
all ``S-describable'' universes computable within countable time
(Def. \ref{quickly}, Section \ref{characterization}); uncountably
many other universes are ignored because they do not even exist
from a constructive point of view.  Postulate \ref{postulate} then
introduces a natural resource-oriented bias reflecting constraints
of whoever calculated our universe (possibly as a by-product of a
search for something else): we assign to universes prior probabilities
inversely proportional to the time and space resources consumed by
the most efficient way of computing them.  Given the resulting ``speed
prior $S$'' (Def. \ref{speedprior}) and past observations $x$, Theorem
\ref{probablyfast} and Corollary \ref{probablyKt} demonstrate that the
best way of predicting a future $y$ is to minimize the Levin complexity
of $(x,y)$.

Section \ref{consequences} puts into perspective the
algorithmic priors (recursive and enumerable) introduced in
previous work on inductive inference by Solomonoff and others
\cite{Solomonoff:64,Solomonoff:78,LiVitanyi:97,Hutter:99}, as well
as the novel priors discussed in the present paper (cumulatively
enumerable, approximable, resource-optimal). Collectively they yield an
entire spectrum of algorithmic TOEs.  We evaluate 
the plausibility of each prior 
being the one from which our own universe is sampled, discuss
its connection to ``Occam's razor'' as well as certain physical and
philosophical consequences, argue that the resource-optimal speed
prior $S$ may be the most plausible one (Section \ref{Splausibility}),
analyze the inference problem from the point of view of an observer
\cite{Boskovich:1755,Boskovich:1966,Toffoli:79,Zurek:91,Svozil:94,Roessler:98}
evolving in a universe sampled from $S$, make appropriate predictions
for our own universe (Section \ref{Spredictions}), and discuss their
falsifiability.

\section{Preliminaries }
\label{preliminaries}

\subsection{Notation} 
Much but not all of the notation used here is similar
or identical to the one used in the standard textbook on Kolmogorov
complexity by Li and Vit\'{a}nyi \cite{LiVitanyi:97}.  

Since sentences
over any finite alphabet are encodable as bitstrings, without loss of
generality we focus on the binary alphabet $B=\{0,1\}$.
$\lambda$ denotes the empty string,
$B^*$ the set of finite sequences over $B$,
$B^{\infty}$ the set of infinite sequences over $B$,
$B^{\sharp} =  B^* \cup B^{\infty}$.
$x,y,z,z^1,z^2$ stand for strings in $B^{\sharp}$.
If $x \in B^*$ then $xy$ is the concatenation of $x$ and $y$ (e.g.,
if $x=10000$ and $y=1111$ then $xy = 100001111$).
Let us order $B^{\sharp}$ lexicographically: if 
$x$ precedes $y$ alphabetically (like in the example above)
then we write $x \prec y$ or $y \succ x$; if
$x$ may also equal $y$ then we write $x \preceq y$ or $y \succeq x$ 
(e.g., $ \lambda \prec 001 \prec 010 \prec 1 \prec 1111... $).
The context will make clear where we also identify $x \in B^*$ with 
a unique nonnegative integer $1x$ 
(e.g., string 0100 is represented by integer 10100 in the dyadic 
system or $20=1*2^4+0*2^3+1*2^2+0*2^1+0*2^0$ in the decimal system).
Indices $i,j,m,m_0,m_1,n,n_0,t,t_0$ range over the positive integers,  
constants $c,c_0,c_1$ over the positive reals,  
$f,g$ denote functions mapping integers to integers,
$log$ the logarithm with basis 2,
$lg(r) = max_k \{integer~k: 2^k \leq r \}$ for real $r>0$.
For $x \in B^* \backslash \{ \lambda \}$, 
$0.x$ stands for the real number with dyadic expansion $x$
(note that $0.x0111....=0.x1=0.x10=0.x100...$ for $x \in B^*$,
although $x0111....\neq x1 \neq x10 \neq x100...$).
For $x \in B^*$, $l(x)$ denotes the number of bits in $x$,
where 
$l(x)= \infty$ for $x \in B^{\infty}$; $l(\lambda) = 0$. 
$x_{n}$ is the prefix of $x$ consisting of 
the first $n$ bits, if $l(x) \geq n$,
and $x$ otherwise ($x_0 := \lambda$). 
For those $x \in B^*$ that contain at least one 0-bit, 
$x'$ denotes the lexicographically smallest $y \succ x$
satisfying $l(y) \leq l(x)$ ($x'$ is undefined for $x$ of the form $111\ldots 1$).
We write $f(n)=O(g(n))$ if there
exists $c,n_0$ such that $f(n) \leq cg(n)$ for all $n > n_0$.

\subsection{Turing Machines: Monotone TMs (MTMs), 
General TMs (GTMs), Enumerable Output Machines (EOMs) 
} 

The standard model of theoretical computer science
is the Turing Machine (TM).  It allows for emulating any known computer.
For technical reasons we will consider several types of TMs.

{\bf Monotone TMs (MTMs).} Most current theory of
description size and inductive inference is based on MTMs 
(compare \cite[p. 276 ff]{LiVitanyi:97})
with several tapes, each tape being
a finite chain of adjacent squares with a scanning
head initially pointing to the leftmost square.  
There is one output tape and at least two work tapes 
(sufficient to compute
everything traditionally regarded as computable).  
The MTM has a finite
number of internal states, one of them being the initial state. MTM
behavior is specified by a lookup table mapping current state and
contents of the squares above work tape scanning heads to a new state
and an instruction to be executed next.  There are instructions for
shifting work tape scanning heads one square left or right (appending
new squares when necessary), and for writing 0 or 1 on squares above
work tape scanning heads. The only input-related instruction requests
an input bit determined by an external process and copies it onto the
square above the first work tape scanning head.  There may or may not
be a halt instruction to terminate a computation. Sequences of requested
input bits are called {\em self-delimiting programs} because they convey
all information about their own length, possibly causing the MTM to halt
\cite{Levin:74,Gacs:74,Chaitin:75}, or at least to cease requesting new input
bits (the typical case in this paper).  MTMs are called {\em monotone}
because they have a one-way write-only
output tape --- they cannot edit their previous output, because
the only ouput instructions are: append a new square 
at the right end of the output tape and fill it with 0/1.

{\bf General TMs (GTMs).} GTMs are like MTMs but have
additional output instructions to edit their previous output.
Our motivation for introducing GTMs is
that certain bitstrings are compactly describable on nonhalting GTMs but 
not on MTMs, as will be seen later.
This has consequences for 
definitions of individual describability and 
probability distributions on describable things.
The additional instructions are: (a) shift output scanning head right/left
(but not out of bounds); (b) delete square at the right end of the
output tape (if it is not the initial square or above the scanning head);
(c) write 1 or 0 on square above output scanning head.
Compare Burgin's inductive TMs and super-recursive 
algorithms \cite{Burgin:83,Burgin:91}.

{\bf Enumerable Output Machines (EOMs).} 
Like GTMs, EOMs can edit their previous output, but not such that it
decreases lexicographically.  The expressive power of EOMs lies
in between those of MTMs and GTMs, with interesting computability-related
properties whose analogues do not hold for GTMs.  EOMs are like MTMs,
except that the only permitted output instruction sequences are:
(a) shift output tape scanning head left/right unless this leads
out of bounds; (b) replace bitstring starting above the output
scanning head by the string to the right of the scanning head of the
second work tape, readjusting output tape size accordingly, but only
if this lexicographically increases the contents of the output tape.
The necessary test can be hardwired into the finite TM transition table.

\subsection{Infinite Computations, Convergence, Formal Describability} 

Most traditional computability theory focuses on
properties of halting programs.
Given an MTM or EOM or GTM $T$ with halt instruction
and $p,x \in B^*$,
we write 
\begin{equation}
T(p)=x
\end{equation}
for ``$p$ computes $x$ on $T$ and halts''.
Much of this paper, however, deals with programs 
that never halt, and with TMs that do not need halt instructions.

\begin{definition}[Convergence] 
\label{convergence}
Let $p \in B^{\sharp}$ denote
the input string or program read by TM T.  Let $T_t(p)$ denote T's finite output
string after $t$ instructions.  We say that $p$ and $p$'s output {\em
stabilize} and {\em converge} towards $x \in B^{\sharp}$ iff for each
$n$ satisfying $0 \leq n \leq  l(x)$ there exists a postive integer
$t_n$ such that for all $t \geq t_n$: 
$T_t(p)_{n} = x_{n}$ and $l(T_t(p)) \leq  l(x)$.
Then we write 
\begin{equation}
T(p) \leadsto x.
\end{equation}
\end{definition}
Although each beginning or prefix of $x$ eventually becomes stable during the possibly
infinite computation,  there need not be a halting program that computes
an upper bound of stabilization time, given any $p$ and prefix size.
Compare the concept of computability {\em in the limit}
\cite{Gold:65,Putnam:65,Freyvald:74} and \cite{Greg:57,Mostowski:57}.

\begin{definition}[TM-Specific Individual Describability] 
\label{TMdescribability}
Given a TM T,
an $x \in B^{\sharp}$ is {\em T-describable} or {\em T-computable} iff
there is a  finite $p \in B^*$ such that $T(p) \leadsto x$.
\end{definition}
Objects with infinite shortest descriptions on $T$ are not $T$-describable.

\begin{definition}[Universal TMs] 
\label{universalTMs}
Let $C$ denote a set of TMs.  $C$ has a
universal element if there is a TM $U^C \in C$ such that for each $T \in C$
there exists a constant string $p_T \in B^*$ (the {\em compiler}) such
that for all possible programs $p$, if $T(p) \leadsto x$ then $U^C(p_T p)
\leadsto x$.
\end{definition}

\begin{definition} [M, E, G] 
\label{MEG}
Let $M$ denote the set of MTMs, $E$ denote the set of EOMs, $G$ denote the
set of GTMs. 
\end{definition}
$M,E,G$ all have universal elements, according to
the fundamental {\em compiler theorem} (for instance, a fixed compiler 
can translate arbitrary LISP programs into equivalent FORTRAN programs).

\begin{definition}[Individual Describability] 
\label{individualdescribability}
Let $C$ denote a set
of TMs with universal element $U^C$.  Some $x \in B^{\sharp}$ is {\em
C-describable} or {\em C-computable} if it is $U^C$-describable.  
E-describable strings are called {\em enumerable}.  G-describable strings
are called {\em formally describable} or simply {\em describable}.
\label{describability}
\end{definition}

\begin{example}[Pseudorandom universe based on halting problem]
\label{universe2}
{\em
Let $x$ be a universe history in the style of Example \ref{universe}.
Suppose its pseudorandom generator's  $n$-th output bit $PRG(n)$ is 1 if
the $n$-th program of an ordered list of all possible programs halts, and
0 otherwise. Since $PRG(n)$ is describable, $x$ is too.  But there is no
halting algorithm computing $PRG(n)$ for all $n$, otherwise the
halting problem would be solvable, which it is not \cite{Turing:36}.
Hence in general there is no computer that outputs $x$ {\em and only} $x$
without ever editing some previously computed history.  
} 
\end{example}

\begin{definition}[Always converging TMs] 
\label{convergingTMs}
TM $T$ 
always converges if for all of its possible 
programs $p \in B^{\sharp}$ there is an
$x \in B^{\sharp}$ such that $T(p) \leadsto x$.
\end{definition}
For example, MTMs and EOMs converge always. GTMs do not.

\begin{definition}[Approximability] 
\label{approximability}
Let $0.x$ denote a real
number,  $x \in B^{\sharp} \backslash \{ \lambda \}$.  
$0.x$ is {\em approximable} by TM $T$ if
there is a $p \in B^*$ such that for each real $\epsilon > 0$ there
exists a $t_0$ such that 
\[
\mid 0.x - 0.T_t(p) \mid < \epsilon
\]
for all
times $t \geq t_0$.  $0.x$ is {\em approximable} if there is at least
one GTM  $T$ as above --- compare (\ref{approx}).
\end{definition}

\begin{lemma}
\label{describabilityapproximability}
If $0.x$ is approximable, then $x$ is describable, and vice versa.  
\end{lemma}

\subsection{Formally Describable Functions} 

Much of the traditional theory of computable functions focuses
on halting programs that map subsets of $B^*$ to subsets of $B^*$. 
The output of a program that does not halt is usually
regarded as undefined, which is occasionally expressed by 
notation such as $T(p) = \infty$.
In this paper, however, we will not lump together
all the possible outputs of nonhalting programs
onto a single symbol ``undefined.'' Instead we will
consider mappings from subsets of $B^*$ to subsets 
of $B^{\sharp}$, sometimes from
$B^{\sharp}$ to $B^{\sharp}$.

\begin{definition}[Encoding $B^*$]
\label{encoding}
Encode $x \in B^*$ as a self-delimiting input $p(x)$
for an appropriate TM, using 
\begin{equation}
l(p(x)) = l(x) + 2log~l(x) + O(1)
\end{equation}
bits as follows: write $l(x)$ in binary notation, 
insert a ``0'' after every ``0'' and a ``1'' after every ``1,'' 
append ``01'' to indicate the
end of the description of the size of the following string,
then append $x$. 
\end{definition}
For instance, $x=01101$ gets encoded as $p(x)=1100110101101$.

\begin{definition}[Recursive Functions]
\label{recursivefns}
A function $h: D_1 \subset B^* \rightarrow D_2 \subset B^*$ is
recursive if there is a TM $T$ using the encoding \ref{encoding} such that
for all $x \in D_1: T(p(x)) = h(x)$.  
\end{definition}

\begin{definition}[Describable Functions]
\label{describablefns}
Let $T$ denote a TM using the encoding of Def. \ref{encoding}.
A function $h: D_1 \subset B^* \rightarrow D_2 \subset B^{\sharp}$ is
$T$-describable if 
for all $x \in D_1: T(p(x)) \leadsto h(x)$.  
Let $C$ denote a set of TMs using encoding \ref{encoding},
with universal element $U^C$.  $h$ is {\em C-describable}
or {\em C-computable} if it is {\em $U^C$-computable.} 
If the $T$ above is universal 
among the GTMs with such input encoding
(see Def. \ref{universalTMs})
then $h$ is {\em describable}.
\end{definition}
Compare functions in the {\em arithmetic hierarchy} \cite{Rogers:67}
and the concept of $\Delta^0_n$-describability, e.g., 
\cite[p. 46-47]{LiVitanyi:97}.

\subsection{Weak Decidability and Convergence Problem} 

Traditionally, decidability of some problem class implies there is a
halting algorithm that prints out the answer, given a problem from
the class.  We now relax the notion of decidability by allowing for
infinite computations on EOMs or GTMs whose answers
converge after finite yet possibly unpredictable time.
Essentially, an answer needs to be correct for almost all the time,
and may be incorrect for at most finitely many initial time steps
(compare computability in the limit \cite{Greg:57,Gold:65,Putnam:65,Freyvald:74}
and super-recursive algorithms \cite{Burgin:83,Burgin:91}).

\begin{definition}[Weak decidability]
\label{weak}
Consider a {\em characteristic function} $h:
D_1 \subset B^* \rightarrow B$: $h(x)=1$ if $x$ satisfies a certain
property, and $h(x)=0$ otherwise.  The problem of deciding whether or not
some $x \in D_1 $ satisfies that property is {\em weakly decidable} if
$h(x)$ is describable (compare Def. \ref{describablefns}).
\end{definition}

\begin{example}
{\em
Is a given string $p \in B^*$ a halting program
for a given MTM?  The problem is not decidable in the traditional sense
(no halting algorithm solves the general halting problem \cite{Turing:36}), 
but weakly
decidable and even E-decidable, by a trivial algorithm: print ``0'' on
first output square; simulate the MTM on work tapes and apply it to $p$,
once it halts after having read no more than $l(p)$ bits 
print ``1'' on first output square.
}
\end{example}

\begin{example}
{\em
It is weakly decidable whether a finite bitstring $p$
is a program for a given TM. Algorithm: print ``0''; feed $p$ bitwise into
the internally simulated TM whenever it requests a new input bit; once the
TM has requested $l(p)$ bits, print ``1''; if it requests an additional
bit, print ``0''.  After finite time the output will stabilize forever.
}
\end{example}

\begin{theorem}[Convergence Problem]
Given a GTM, it is not weakly decidable whether a
finite bitstring is a converging program, or whether 
some of the output bits will fluctuate forever.
\label{fluctuate}
\end{theorem}

\noindent {\bf Proof.} 
A proof conceptually quite similar to the one below
was given by Hotz, Vierke and Schieffer  \cite{Hotz:95}
in the context of analytic TMs \cite{Hotz:99} 
derived from R-Machines \cite{Blum:89} (the
alphabet of analytic TMs is {\em real-valued} instead of binary).  Version
1.0 of this paper \cite{Schmidhuber:00version1} was written without
awareness of this work. Nevertheless, the proof in Version 1.0 is
repeated here because it does serve illustrative purposes.

In a straightforward manner we adapt Turing's proof of the 
undecidability of the MTM halting problem \cite{Turing:36}, 
a reformulation of G\"{o}del's celebrated result \cite{Goedel:31},
using the diagonalization trick whose roots date back to Cantor's proof
that one cannot count the real numbers \cite{Cantor:1874}.  
Let us write $T(x) \downarrow$
if there is a $z \in B^{\sharp}$ such that $T(x) \leadsto z$. 
Let us write $T(x) \updownarrow$
if $T$'s output fluctuates
forever in response to $x$ (e.g., by flipping from 1 to zero
and back forever).  
Let $A_1, A_2, \ldots$ be an effective
enumeration of all GTMs.  Uniquely encode all pairs of finite strings
$(x,y)$ in $B^* \times B^*$ as finite strings $code(x,y) \in B^*$.
Suppose there were
a GTM U such that {\bf (*)}: for all $x,y \in B^*$ : $U(code(x,y))
\leadsto 1$ if  $A_x(y) \downarrow$, 
and $U(code(x,y)) \leadsto 0$ otherwise.
Then one could construct a GTM $T$ with 
$T(x) \leadsto 1$ 
if $U(code(x,x)) \leadsto 0$, and $T(x) \updownarrow$ otherwise.
Let $y$ be the index of $T = A_y$, then 
$A_y(y) \downarrow$ if $U(code(y,y)) \leadsto 0$,  
otherwise $A_y(y) \updownarrow$.  
By {\bf (*)}, however, 
$U(code(y,y)) \leadsto 1$ if $A_y(y) \downarrow$, and  
$U(code(y,y)) \leadsto 0$ if $A_y(y) \updownarrow$.  
Contradiction. $\Box$

\section{Complexity of Constructive Descriptions}
\label{descriptive} 

Throughout this paper we focus on
TMs with self-delimiting programs \cite{Levin:73a,Levin:74,Gacs:74,Chaitin:75}.
Traditionally, the Kolmogorov complexity 
\cite{Kolmogorov:65,Solomonoff:64,Chaitin:69} 
or algorithmic complexity or algorithmic information
of $x \in B^*$ is
the length of the shortest halting program computing $x$: 

\begin{definition}[Kolmogorov Complexity $K$]
Fix a universal MTM or EOM or GTM U with halt instruction, 
and define
\begin{equation}
K(x) = \min_{p}\{l(p) : U(p)=x \}.
\label{K}
\end{equation}
\end{definition}
Let us now extend this to nonhalting GTMs.

\subsection{Generalized Kolmogorov Complexity for EOMs and GTMs}

\begin{definition}[Generalized $K_T$]
\label{generalizedK}
Given any TM T, define
\[
K_T(x) = \min_p\{l(p): T(p) \leadsto x \}
\]
\end{definition}
Compare Schnorr's ``process complexity'' for MTMs \cite{Schnorr:73,Vyugin:98}. 

\begin{definition}[$K^M, K^E, K^G$ based on Invariance Theorem]
\label{invariance}
Consider Def. \ref{MEG}.
Let $C$ denote a set of TMs with universal TM $U^C$  ($T \in C$).
We drop the index $T$, writing 
\[
K^C(x) = K_{U^C}(x) \leq  K_T(x) + O(1).
\]
\end{definition}
This is justified by an appropriate {\em Invariance Theorem} 
\cite{Kolmogorov:65,Solomonoff:64,Chaitin:69}:
there is a positive constant $c$ such
that $K_{U^C}(x) \leq  K_T(x) + c $ for all $x$,
since the size  of the compiler that translates arbitrary
programs for $T$ into equivalent programs for $U^C$ does not depend on $x$.

\begin{definition}[$Km_T,Km^M,Km^E,Km^G$]
\label{generalizedKm}
Given TM $T$ and $x \in B^*$, define
\begin{equation}
Km_T(x) = \min_p\{l(p) : T(p) \leadsto xy, y \in B^{\sharp} \}.
\label{Km}
\end{equation}
Consider Def. \ref{MEG}.
If $C$ denotes a set of TMs with universal TM $U^C$, then define
$Km^C(x) = Km_{U^C}(x).$ 
\end{definition}
$Km^C$ is a generalization of Schnorr's \cite{Schnorr:73} 
and Levin's \cite{Levin:73a} complexity measure  $Km^M$ for MTMs. 

\vspace{0.3cm} \noindent
 {\bf Describability issues.}
$K(x)$ is not computable by a halting program
\cite{Kolmogorov:65,Solomonoff:64,Chaitin:69}, but obviously 
$G$-computable or describable; the $z$ with $0.z =$ $1 \over K(x)$ 
is even enumerable.  Even $K^E(x)$ is describable, using the following algorithm: 

\begin{quote}
Run all EOM programs in {\em ``dovetail style''} such
that the $n$-th step of the $i$-th program is executed in the $n+i$-th
phase ($i = 1, 2, \ldots$); whenever a program outputs $x$, place it
(or its prefix read so far) in a tentative list $L$ of $x$-computing
programs or program prefixes; whenever an element of $L$ produces output
$\succ x$, delete it from $L$; whenever an element of $L$ requests an
additional input bit, update $L$ accordingly.  After every change of
$L$ replace the current estimate of $K^E(x)$ by the length of the 
shortest element of $L$. This estimate will eventually stabilize forever.
\end{quote}

\begin{theorem}
$K^G(x)$ is not describable.
\label{KGnot}
\end{theorem}

\noindent {\bf Proof.} Identify finite bitstrings with the integers they represent.
If $K^G(x)$ were describable then also 
\begin{equation}
h(x)= max_y \{K^G(y): 1 \leq y \leq g(x) \},
\end{equation}
where $g$ is any fixed recursive function, and also
\begin{equation}
f(x)= min_y \{y: K^G(y) = h(x) \}.
\label{f}
\end{equation}
Since the number of descriptions $p$ with $l(p) < n-O(1)$ 
cannot exceed $2^{n-O(1)}$, but the number of strings $x$ with $l(x) = n$ equals $2^n$,
most $x$ cannot be compressed by more than $O(1)$ bits; that is, 
$K^G(x) \geq   log~x - O(1)$ for most $x$.
From (\ref{f}) we 
therefore obtain $K^G(f(x)) >  log~g(x) - O(1)$ for large enough $x$, because $f(x)$
picks out one of the incompressible $y \leq g(x)$.
However, obviously we also would have $K^G(f(x)) \leq l(x) + 2log~l(x) + O(1)$, 
using the encoding of Def. \ref{encoding}.
Contradiction for quickly growing $g$ with low complexity, such as $g(x)=2^{2^x}$. 
$\Box$

\subsection{Expressiveness of EOMs and GTMs}
\label{power}

On their internal work tapes MTMs can compute whatever GTMs can
compute. But they commit themselves forever once they print out some
bit. They are ill-suited to the case where the output may require
subsequent revision after time intervals unpredictable in advance
--- compare Example \ref{universe2}.  Alternative MTMs that print out
sequences of result updates (separated by, say, commas) would compute
other things besides the result, and hence not satisfy the {\em ``don't
compute anything else''} aspect of individual describability.  Recall from
the introduction that in a certain sense there are uncountably many
collectively describable strings, but only countably many individually
describable ones.

Since GTMs may occasionally rewrite parts of their output,
they are computationally more expressive than MTMs  in the sense
that they permit much more compact descriptions of certain
objects. 
For instance, 
$K(x) - K^G(x)$ 
is unbounded, as will be seen next. 
This will later have 
consequences for predictions, given certain observations.

\begin{theorem}
$K(x) - K^G(x)$ is unbounded.
\label{K-KG}
\end{theorem}
{\bf Proof.}
Define
\begin{equation}
h'(x)= max_y \{K(y): 1 \leq y \leq g(x) \};~~ f'(x)= min_y \{y: K(y) = h'(x) \},
\end{equation}
where $g$ is recursive. 
Then $K^G(f'(x)) = O(l(x) + K(g))$ (where $K(g)$ is the size of the minimal
halting description of function $g$), but $K(f'(x)) > log~g(x) -O(1)$ for 
sufficiently large $x$ ---
compare the proof of Theorem \ref{KGnot}.
Therefore $K(f'(x)) - K^G(f'(x)) \geq O(log~g(x))$ for infinitely many
$x$ and quickly growing $g$ with low complexity. 
$\Box$

\subsubsection{EOMs More Expressive Than MTMs} 

Similarly, some $x$ are compactly describable on EOMs 
but not on MTMs.  To see this,
consider Chaitin's $\Omega$, the halting probability of an MTM
whose input bits are obtained by tossing an unbiased coin whenever it
requests a new bit \cite{Chaitin:87}.  $\Omega$ is enumerable (dovetail over
all programs $p$ and sum up the contributions $2^{-l(p)}$ of the halting $p$), but 
there is no recursive upper bound on the number of
instructions required to compute $\Omega_{n}$, given $n$.  This implies
$K(\Omega_{n}) = n + O(1)$  \cite{Chaitin:87} and also
$K^M(\Omega_{n}) = n + O(1)$.  It is easy to see, however,
that on nonhalting EOMs there are much more compact descriptions: 
\begin{equation}
K^E(\Omega_{n}) \leq O(K(n)) \leq O(log~n);  
\end{equation}
that is, there is no upper bound of
\begin{equation}
K^M(\Omega_{n}) - K^E(\Omega_{n}). 
\label{KE-KM}
\end{equation}

\subsubsection{GTMs More Expressive Than EOMs --- Objects Less Regular Than $\Omega$}

We will now show that there are describable strings that have a
short GTM description yet are ``even more random'' than Chaitin's Omegas,
in the sense that even on EOMs they do not have any compact descriptions.

\begin{theorem} For all $n$ there are $z \in B^*$ with 
\[
K^E(z) > n - O(1),~~yet~~ K^G(z) \leq O(log~n).
\]
That is, $K^E(z) - K^G(z)$ is unbounded.
\label{z}
\end{theorem}

 \noindent {\bf Proof.} For $x \in B^* \backslash \{ \lambda \}$ and universal
EOM $T$ define 
\begin{equation}
\Xi(x) = 
\sum_{y \in B^{\sharp}: 0.y > 0.x} ~~
\sum_{p: T(p) \leadsto y} 2^{-l(p)}.
\end{equation}
First note that the dyadic expansion of 
$\Xi(x)$ is EOM-computable or enumerable. 
The algorithm works as follows:
\begin{quote}
Algorithm A: Initialize the real-valued variable $V$ by 0, run all possible programs of
EOM $T$ dovetail style such that the $n$-th step of the $i$-th program 
is executed in the $n+i$-th phase;
whenever the output of a program prefix $q$ starts
with some $y$ satisfying $0.y > 0.x$ for the first time, set $V:= V+
2^{-l(q)}$; henceforth ignore continuations of $q$.  
\end{quote}
$V$ approximates $\Xi(x)$ from below in enumerable fashion --- infinite $p$
are not worrisome as $T$ must only read a finite prefix of $p$ to observe
$0.y > 0.x$ if the latter holds indeed.
We will now show that knowledge of $\Xi(x)_n$,
the first $n$ bits of $\Xi(x)$,
allows for constructing a bitstring $z$ with 
$K^E(z) \geq n - O(1)$ when $x$ has low complexity.

Suppose we know $\Xi(x)_n$. 
Once algorithm A above yields $V >
\Xi(x)_n$ we know that no programs $p$ with $l(p) < n$ will contribute
any more to V.  Choose the shortest $z$ satisfying $0.z = (0.y_{min} - 0.x)/2$,
where $y_{min}$ is the lexicographically smallest 
$y$ previously computed by algorithm A 
such that $0.y > 0.x$.  Then $z$ cannot be among the strings T-describable
with fewer than $n$ bits. Using the Invariance Theorem 
(compare Def. \ref{invariance})
we obtain $K^E(z) \geq  n - O(1)$.

While prefixes of $\Omega$ are greatly compressible on EOMs, 
$z$ is not.
On the other hand, $z$ is compactly G-describable: $K^G(z) \leq K(x) +
K(n) + O(1)$.  For instance, choosing a low-complexity $x$, we have
$K^G(z) \leq O(K(n)) \leq O(log~n)$.  $\Box$

\vspace{0.3cm}
\noindent
The discussion above reveils a natural complexity hierarchy.
Ignoring additive constants, we have
\begin{equation}
\label{hierarchy}
K^G(x) \leq K^E(x) \leq K^M(x),
\end{equation}
where for each ``$\leq$'' relation above 
there are $x$ which allow for replacing
``$\leq$'' by ``$<$.''

\section{Measures and Probability Distributions}
\label{measures}

Suppose $x$ represents the history of our universe up until now.  
What is its most likely continuation $y \in B^{\sharp}$? Bayes' theorem yields
\begin{equation}
P(xy \mid x) 
= \frac{P(x \mid xy) P(xy)} {\sum_{z \in B^{\sharp}}  P(xz)} 
= \frac{P(xy)} {N(x)}  
\propto P(xy) 
\label{bayes2}
\end{equation}
where $P(z^2 \mid z^1)$ is the probability of $z^2$, given knowledge of $z^1$, and
\begin{equation}
N(x) = \sum_{z \in B^{\sharp}}  P(xz)
\end{equation}
is a normalizing factor.  The most likely continuation $y$ is
determined by $P(xy)$, the {\em prior probability} of $xy$ 
--- compare the similar Equation (\ref{bayes1}).
Now what are the formally describable ways of assigning prior probabilities
to universes?
In what follows we will first consider describable {\em semimeasures} on $B^*$,
then probability distributions on $B^{\sharp}$.

\subsection{Dominant and Universal (Semi)Measures }

The next three definitions concerning semimeasures on $B^*$ are almost
but not quite identical to those of discrete semimeasures
\cite[p. 245 ff]{LiVitanyi:97} 
and continuous semimeasures
\cite[p. 272 ff]{LiVitanyi:97} 
based on the work of Levin and Zvonkin
\cite{Zvonkin:70}.

\begin{definition}[Semimeasures]
\label{mu}
A (binary) semimeasure $\mu$ is a function $B^* \rightarrow [0,1]$ that satisfies:
\begin{equation}
\mu(\lambda) = 1;~~ 
\mu(x) \geq 0;~~ 
\mu(x) = \mu(x0) + \mu(x1) + \bar{\mu}(x),
\end{equation}
where 
$\bar{\mu}$ is a function $B^* \rightarrow [0,1]$ satisfying
$0 \leq \bar{\mu}(x) \leq \mu(x)$. 
\end{definition}
A notational difference to the approach of Levin 
\cite{Zvonkin:70}
(who writes $\mu(x) \leq \mu(x0) + \mu(x1)$) 
is the explicit introduction of $\bar{\mu}$. Compare the
introduction of an undefined element $u$ by Li and
Vitanyi \cite[p. 281]{LiVitanyi:97}.
Note that $\sum_{x \in B^*} \bar{\mu}(x) \leq 1$.
Later we will discuss the interesting case $\bar{\mu}(x)=P(x)$, 
the a priori probability of $x$.

\begin{definition}[Dominant Semimeasures]
A semimeasure $\mu_0$ dominates another semimeasure $\mu$ if
there is a constant $c_{\mu}$ such that for all $x \in B^*$
\begin{equation}
\mu_0(x) > c_{\mu} \mu(x).
\label{dominantmeasures}
\end{equation}
\end{definition}

\begin{definition}[Universal Semimeasures]
Let $\cal M$ be a set of semimeasures on $B^*$.
A semimeasure $\mu_0 \in$ $\cal M$ is universal if it
dominates all $\mu \in$ $\cal M$. 
\end{definition}

In what follows,
we will introduce describable semimeasures dominating those considered
in previous work (\cite{Zvonkin:70}, \cite[p. 245 ff, p.272 ff]{LiVitanyi:97}).

\subsection{Universal Cumulatively Enumerable Measure (CEM)}

\begin{definition}[Cumulative measure $C\mu$]
For semimeasure $\mu$ on $B^*$ define
the cumulative measure $C\mu$:
\begin{equation}
C\mu(x) := 
\sum_{y \succeq x:~ l(y)=l(x)} \mu(y) 
+ \sum_{y \succ x:~ l(y) < l(x)} \bar{\mu}(y).
\end{equation}
\end{definition}
Note that we could replace ``$l(x)$'' by ``$l(x) + c$'' 
in the definition above.
Recall that $x'$ denotes the smallest $y \succ x$
with $l(y) \leq l(x)$ ($x'$ may be undefined).
We have
\begin{equation}
\mu(x) = C\mu(x) ~if~x=11...1; ~~else~~
\mu(x) = C\mu(x) - C\mu(x').
\label{Cmu}
\end{equation}

\begin{definition}[CEMs]
Semimeasure $\mu$ is a CEM if 
$C\mu(x)$ is enumerable for all $x \in B^*$.
\label{cem}
\end{definition}
Then $\mu(x)$ is the difference of two finite enumerable values,
according to (\ref{Cmu}).

\begin{theorem}
There is a  universal CEM.
\label{universalCEM}
\end{theorem}

\noindent {\bf Proof.} 
We first show that one can enumerate the CEMs, then construct
a universal CEM from the enumeration. Check out differences
to Levin's related proofs that
there is a universal discrete semimeasure and a
universal enumerable semimeasure \cite{Zvonkin:70,Levin:73a},
and Li and Vit\'{a}nyi's presentation
of the latter \cite[p. 273 ff]{LiVitanyi:97}, 
attributed to J. Tyszkiewicz.

Without loss of generality, consider only EOMs without
halt instruction and with fixed input
encoding of $B^*$ according to
Def. \ref{encoding}.
Such EOMs are enumerable, and correspond
to an effective enumeration of all enumerable functions from $B^*$ to
$B^{\sharp}$. Let $EOM_i$ denote the $i$-th EOM in the list, and let
$EOM_i(x,n)$ denote its output
after $n$ instructions when applied to $x \in B^*$. 
The following procedure filters out those $EOM_i$ that already represent
CEMs, and transforms the others into representations
of CEMs, such that we obtain a way of generating all and only CEMs.

\begin{quote}

{\tt FOR} all $i$ {\tt DO} in dovetail fashion:

\begin{quote}
{\tt START:} 
let 
$V\mu_i(x)$ 
and 
$V\bar{\mu}_i(x)$ 
and
$VC\mu_i(x)$ 
denote variable functions on $B^*$. Set
$V\mu_i(\lambda):= V\bar{\mu}_i(\lambda) := VC\mu_i(\lambda):= 1$,
and 
$V\mu_i(x):= V\bar{\mu}_i(x) := VC\mu_i(x):= 0$ 
for all other $x \in B^*$.
Define $VC\mu_i(x') := 0$ for
undefined $x'$. 
Let $z$ denote a string variable.

{\tt FOR} $n = 1, 2, \ldots$ {\tt DO}:

\begin{quote}

{\bf (1)} 
Lexicographically order and rename all $x$ with $l(x) \leq n$: 

$x^1  := \lambda 
\prec x^2 := 0 \prec x^3  \prec \ldots \prec x^{2^{n+1}-1} := \underbrace {11...1}_n$. 

{\bf (2)}
{\tt FOR } $k = 2^{n+1}-1$ down to 1 {\tt DO}:

\begin{quote}

{\bf (2.1)}
Systematically search for the smallest $m \geq n$ such that 
$z :=EOM_i(x^k,m) \neq  \lambda$ {\tt AND}
$0.z \geq  VC\mu_i(x^{k+1})$ if $k <2^{n+1}-1$;
set $VC\mu_i(x^k) := 0.z$. 

\end{quote}

{\bf (3)} 
For all $x \succ \lambda$ satisfying $l(x) \leq n$,
set $V\mu_i(x) := VC\mu_i(x)  - VC\mu_i(x')$.
For all $x$ with $l(x) < n$,
set $V\bar{\mu}_i(x) := V\mu_i(x) - V\mu_i(x1) - V\mu_i(x0)$. 
For all $x$ with $l(x) = n$, set $V\bar{\mu}_i(x) := V\mu_i(x)$. 

\end{quote}

\end{quote}
\end{quote}
If $EOM_i$ indeed represents a CEM $\mu_i$ then 
each search process in {\bf (2.1)} will terminate, and
the $VC\mu_i(x)$ 
will enumerate the
$C\mu_i(x)$ from below, and 
the $V\mu_i(x)$
and $V\bar{\mu}_i(x)$ 
will approximate the true $\mu_i(x)$
and $\bar{\mu}_i(x)$, respectively,
not necessarily from below though. 
Otherwise there will be a nonterminating search
at some point, leaving $V\mu_i$ from the previous loop 
as a trivial CEM. Hence we can enumerate all CEMs,
and only those.  Now define (compare \cite{Levin:73a}):
\[
\mu_0(x) = \sum_{n>0} \alpha_n \mu_n(x), ~~
\bar{\mu}_0(x) = \sum_{n>0} \alpha_n \bar{\mu}_n(x), ~~ 
~where~ \alpha_n > 0,~ \sum_n \alpha_n = 1,
\]
and $\alpha_n$ is an enumerable constant, e.g., 
$\alpha_n = \frac{6}{\pi n^2}$
or
$\alpha_n = \frac{1}{n(n+1)}$
(note a slight difference to Levin's classic approach 
which just requests $\sum_n \alpha_n \leq 1$).
Then $\mu_0$ 
dominates every $\mu_n$ by Def. \ref{dominantmeasures}, 
and is a semimeasure according to Def. \ref{mu}:
\begin{equation}
\mu_0(\lambda) = 1;~
\mu_0(x) \geq 0;~
\mu_0(x) = 
     \sum_{n>0} \alpha_n [ \mu_n(x0) + \mu_n(x1) + \bar{\mu}_n(x)] =
\mu_0(x0) + \mu_0(x1) + \bar{\mu}_0(x).
\end{equation}
$\mu_0$ also is a CEM by Def. \ref{cem}, because 
\[
C\mu_0(x) = 
\sum_{y \succeq x:~l(x)=l(y)} ~~ \sum_{n>0} \alpha_n \mu_n(y) ~~ +
\sum_{y \succ x:~l(x)>l(y)} ~~ \sum_{n>0} \alpha_n \bar{\mu}_n(y) 
=
\]
\begin{equation}
\sum_{n>0} \alpha_n 
\left(
\sum_{y \succeq x:~l(x)=l(y)} \mu_n(y) +
\sum_{y \succ x:~l(x)>l(y)}   \bar{\mu}_n(y) 
\right)
=
\sum_{n>0} \alpha_n C\mu_n(x) 
\end{equation}
is enumerable, since $\alpha_n$ and $C\mu_n(x)$ are
(dovetail over all $n$).
That is, $\mu_0(x)$ is approximable as the difference of 
two enumerable finite values, according to Equation (\ref{Cmu}).
$\Box$
\vspace{0.5cm}

\subsection{Approximable and Cumulatively Enumerable Distributions}

To deal with infinite $x$, 
we will now extend the treatment of semimeasures on $B^*$ in the
previous subsection by discussing
probability distributions  on $B^{\sharp}$.

\begin{definition}[Probabilities]
\label{prob}
A probability distribution $P$ on $x \in B^{\sharp}$ satisfies 
\[
P(x) \geq 0; ~~\sum_{x} P(x) = 1.
\]
\end{definition}

\begin{definition}[Semidistributions]
A semidistribution $P$ on $x \in B^{\sharp}$ satisfies 
\[
P(x) \geq 0; ~~\sum_{x} P(x) \leq  1.
\]
\end{definition}

\begin{definition}[Dominant Distributions]
A distribution $P_0$ dominates another distribution $P$ if
there is a constant $c_P > 0$ such that for all $x \in B^{\sharp}$:
\begin{equation}
P_0(x) \geq c_P P(x).
\label{dominance}
\end{equation}
\end{definition}

\begin{definition}[Universal Distributions] 
Let $\cal P$ be a set of probability distributions on $x \in B^{\sharp}$.
A distribution $P_0 \in$ $\cal P$ is universal if for all
$P \in$ $\cal P$: $P_0$ dominates $P$.
\end{definition}

\begin{theorem}
There is no universal approximable semidistribution.
\label{nouniversal}
\end{theorem}

\noindent {\bf Proof.} The following proof is due to M. Hutter
(personal communications by email following a discussion of
enumerable and approximable universes on 2 August 2000  in
Munich). It is an extension of a modified\footnote{ As pointed out
by M. Hutter (14 Nov. 2000, personal communication) and even
earlier by A. Fujiwara (1998, according to P. M. B. Vit\'{a}nyi,
personal communication, 21 Nov. 2000), the proof on the bottom of
p. 249 of \cite{LiVitanyi:97} should be slightly modified. For
instance, the sum could be taken over $x_{i-1}<x\leq x_i$.  The
sequence of inequalities $\sum_{x_{i-1}<x\leq x_i}P(x)>x_iP(x_i)$
is then satisfiable by a suitable $x_i$ sequence, since
$\liminf_{x\to\infty}\{xP(x)\}=0$.  The basic idea of the proof is
correct, of course, and very useful. } proof \cite[p. 249 ff]{LiVitanyi:97} 
that there is no universal recursive semimeasure.

It suffices to focus on $x \in B^*$. Identify strings with
integers, and assume $P(x)$ is a universal approximable
semidistribution.  We construct an approximable semidistribution
$Q(x)$ that is not dominated by $P(x)$, thus contradicting the
assumption. Let $P_0(x), P_1(x), \ldots$ be a sequence of
recursive functions converging to $P(x)$. We recursively define a
sequence $Q_0(x), Q_1(x), \ldots$ converging to $Q(x)$. The basic
idea is: each contribution to $Q$ is the sum of $n$ consecutive
$P$ probabilities ($n$ increasing). Define $Q_0(x):=0$;
$I_n:=\{y:n^2\!\leq\!y\!<\!(n\!+\!1)^2\}$. Let $n$ be such that
$x\!\in\!I_n$. Define $j_t^n$ $(k_t^n)$ as the element with
smallest $P_t$ (largest $Q_{t-1}$) probability in this interval,
i.e., $j_t^n:=\minarg_{x\in I_n}P_t(x)$ $(k_t^n:=\maxarg_{x\in
I_n}Q_{t-1}(x))$. If $n\!\cdot\!P_t(k_t^n)$ is less than twice and
$n\!\cdot\!P_t(j_t^n)$ is more than half of $Q_{t-1}(k_t^n)$, set
$Q_t(x)=Q_{t-1}(x)$. Otherwise set $Q_t(x)=n\!\cdot\!P_t(j_t^n)$
for $x=j_t^n$ and $Q_t(x)=0$ for $x\neq j_t^n$. $Q_t(x)$ is
obviously total recursive and non-negative. Since
$2n\!\leq\!|I_n|$, we have $$ \sum_{x\in I_n}Q_t(x) \leq
2n\!\cdot\!P_t(j_t^n) = 2n\!\cdot\!\min_{x\in I_n}P_t(x) \leq
\sum_{x\in I_n}P_t(x). $$ Summing over $n$ we observe that if
$P_t$ is a semidistribution, so is $Q_t$.  From some $t_0$ on,
$P_t(x)$ changes by less than a factor of 2 since $P_t(x)$
converges to $P(x)\!>\!0$. Hence $Q_t(x)$ remains unchanged for
$t\!\geq\!t_0$ and converges to $Q(x):=Q_\infty(x)=Q_{t_0}(x)$.
But $Q(j_{t_0}^n)=Q_{t_0}(j_{t_0}^n)\geq
n\!\cdot\!P_{t_0}(j_{t_0}^n) \geq{\odt}n\!\cdot\!P(j_{t_0}^n)$,
violating our universality assumption $P(x)\geq c\!\cdot\!Q(x)$.
$\Box$

\begin{definition}[Cumulatively Enumerable Distributions -- CEDs]
\label{defCP}
A distribution $P$ on $B^{\sharp}$ is a CED
if $CP(x)$ is enumerable for all $x \in B^*$, where
\begin{equation}
CP(x) := \sum_{y \in B^{\sharp}: y \succeq x} P(y)
\label{CP}
\end{equation}
\end{definition}

\subsection{TM-Induced Distributions and Convergence Probability}

Suppose TM $T$'s input bits are obtained by tossing an unbiased coin 
whenever a new one is requested.  
Levin's {\em universal discrete enumerable semimeasure} 
\cite{Levin:73a,Chaitin:75,Gacs:74}
or {\em semidistribution} $m$
is limited to $B^*$ and halting programs:
\begin{definition}[m]
\label{m}
\begin{equation}
m(x) = \sum_{p: T(p)=x} 2^{-l(p)}; 
\end{equation}
\end{definition}
Note that $\sum_x m(x) < 1$ if $T$ universal.
Let us now generalize this to $B^{\sharp}$ and nonhalting programs:

\begin{definition}[$P_T, KP_T$]
\label{continua} 
Suppose $T$'s input bits are obtained by tossing an unbiased coin 
whenever a new one is requested.  
\begin{equation}
P_T(x) = \sum_{p: T(p) \leadsto x} 2^{-l(p)},~~
KP_T(x) = -lg P_T(x)~for~P_T(x)>0,
\end{equation}
where $x, p \in B^{\sharp}$.  
\end{definition}
{\bf Program Continua.} 
According to Def. \ref{continua}, most 
infinite $x$ have zero probability, but not those with finite
programs, such as the dyadic expansion of $0.5 \sqrt{2}$.  However,
a nonvanishing part of the entire unit of probability mass is
contributed by continua of mostly incompressible strings, such as those
with cumulative probability $2^{-l(q)}$ computed by the following class of
uncountably many infinite programs with a common finite prefix $q$: {``repeat
forever: read and print next input bit.''}
The corresponding traditional measure-oriented notation for 
\[
\sum_{x: T(qx) \leadsto x} 2^{-l(qx)} = 2^{-l(q)}
\]
would be
\[
\int_{0.q}^{0.q + 2^{-l(q)}}  dx  = 2^{-l(q)}.
\]
For notational simplicity, however, we will continue using the $\sum$
sign to indicate summation over uncountable objects, rather than using
a measure-oriented notation for probability densities. 
The reader should not feel uncomfortable
with this --- the theorems in the remainder of the paper 
will focus on those $x \in B^{\sharp}$ with $P(x)>0$; density-like nonzero sums 
over uncountably many bitsrings, each 
with individual measure zero, will not play any critical role in the proofs.

\begin{definition}[Universal TM-Induced Distributions $P^C;KP^C$]
If $C$ denotes a set of TMs with universal element $U^C$, then we write 
\begin{equation}
P^C(x) = P_{U^C}(x); ~~  
KP^C(x) := -lg~P^C(x) ~for~ P^C(x)>0.
\end{equation}
\end{definition}
We have $P^C(x) > 0$ for $D_C \subset B^{\sharp}$, the subset of C-describable  
$x \in B^{\sharp}$.  The attribute {\em universal} is justified, because 
of the dominance $P_T(x) = O(P^C(x))$, 
due to the Invariance Theorem 
(compare Def. \ref{invariance}).

Since all programs of EOMs and MTMs
converge, $P^E$ and $P^M$ are proper probability distributions on $B^{\sharp}$. 
For instance, $\sum_x P^E(x) = 1$.
$P^G$, however, is just a semidistribution. 
To obtain a proper probability 
distribution $PN_T$, one might think of normalizing 
by the {\em convergence probability} $\Upsilon$:

\begin{definition}[Convergence Probability] 
\label{upsilon}
Given GTM T, define
\[ 
PN_T(x) = \frac{\sum_{T(p) \leadsto x} 2^{-l(p)}}
	      {\Upsilon^T}, 
\]
where
\[ 
\Upsilon^T =  \sum_{p: \exists x: T(p) \leadsto x} 2^{-l(p)}. 
\]
\end{definition}
{\bf Describability issues.}
Uniquely encode each TM $T$ as a finite bitstring, 
and identify $M, E, G$ with the corresponding sets of bitstrings.
While the function $f^M: M \rightarrow B^{\sharp}: f(T) = \Omega^T$ 
is describable, even enumerable,
the function $f^G: G \rightarrow B^{\sharp}: f(T) = \Upsilon^T$ 
is not, essentially due to Theorem \ref{fluctuate}.

Even $P^E(x)$ and $P^M(x)$ are generally not describable for $x \in B^{\sharp}$, in the sense
that there is no GTM $T$ that takes as an input a finite description (or
program) of any M-describable or E-describable $x \in B^{\sharp}$ and
converges towards $P^M(x)$ or $P^E(x)$. This is because in general it is
not even weakly decidable (Def. \ref{weak})
whether two programs compute the same output.
If we know that one of the program outputs is finite, however, then the
conditions of weak decidability are fulfilled. Hence certain TM-induced
distributions on $B^*$ are describable, as will be seen next.

\begin{definition}[TM-Induced Cumulative Distributions]
\label{CPC}
If $C$ denotes a set of TMs with universal element $U^C$, then we write 
(compare Def. \ref{defCP}):
\begin{equation}
CP^C(x) = CP_{U^C}(x).  
\end{equation}
\end{definition}

\begin{lemma}
\label{CPEenumerable}
For $x \in B^*$, $CP^E(x)$ is enumerable.
\end{lemma}

\noindent {\bf Proof.} The following algorithm computes $CP^E(x)$ (compare 
proof of Theorem \ref{z}): 

\begin{quote}
Initialize the
real-valued variable $V$ by 0, run all possible programs of EOM $T$ dovetail
style; whenever the output of a program prefix $q$ starts with some $y \succeq x$
for the first time, set $V:= V+ 2^{-l(q)}$;
henceforth ignore continuations of $q$.  
\end{quote}
In this way $V$ enumerates
$CP^E(x)$. Infinite $p$ are not
problematic as only a finite prefix of $p$ must be read to establish $y \succeq x$ 
if the latter indeed holds.  $\Box$
\vspace{0.5cm}

\noindent
Similarly, facts of the form $y \succ x \in B^*$
can be discovered after finite time.

\begin{corollary}
For $x \in B^*$,
$P^E(x)$ is approximable or describable as the difference of two enumerable values:
\begin{equation}
P^E(x) = \sum_{y \succeq x}  P^E(y) - \sum_{y \succ x}  P^E(y),
\label{twoenumerable}
\end{equation}
\end{corollary}
Now we will make the connection to the previous subsection
on semimeasures on $B^*$.

\subsection{Universal TM-Induced Measures} 

\begin{definition}[P-Induced Measure $\mu P$] 
Given a distribution $P$ on $B^{\sharp}$,
define a measure $\mu P$ on $B^*$ as follows:
\begin{equation}
\mu P(x) = \sum_{z \in B^{\sharp}} P(xz).
\end{equation}
\end{definition}
Note that $\overline{\mu P}(x) = P(x)$ 
(compare Def. \ref{mu}):
\begin{equation}
\mu P(\lambda) = 1; ~~\mu P(x) = P(x) + \mu P(x0) + \mu P(x1).
\end{equation}
For those $x \in B^*$ without 0-bit we have
$ \mu P(x) = CP(x)$, for the others 
\begin{equation}
\mu P(x) = CP(x) - CP(x').
\end{equation}

\begin{definition}[TM-Induced Semimeasures $\mu_T,\mu^M,\mu^E,\mu^G$]
\label{muT}
Given some TM $T$,
for $x \in B^*$ define $\mu_T(x) = \mu P_T(x)$. 
Again we deviate a bit from Levin's $B^*$-oriented path 
\cite{Zvonkin:70} (survey: \cite[p. 245 ff,  p. 272 ff]{LiVitanyi:97}) 
and extend $\mu_T$ to $x \in B^{\infty}$, where we define $\mu_T(x) = \bar{\mu}_T(x) = P_T(x)$.
If $C$ denotes a set of TMs with universal element $U^C$, then we write 
\begin{equation}
\mu^C(x) = \mu_{U^C}(x); ~~  
K\mu^C(x) := -lg~\mu^C(x) ~for~ \mu^C(x)>0.
\end{equation}
\end{definition}
We observe that $\mu^C$ is universal among 
all T-induced semimeasures, $T \in C$.  Note that 
\begin{equation}
\mu^C(x) = \mu^C(x0) + \mu^C(x1) + P^C(x) ~for~x \in B^*; ~~
\mu^C(x) = P^C(x) ~for~x \in B^{\infty}.
\end{equation}
It will be obvious from the context when we deal with the
restriction of $\mu^C$ to $B^*$.

\begin{corollary}
For $x \in B^*$,
$\mu^E(x)$  is a CEM and approximable as the difference of two enumerable values:
$ \mu^E(x) = CP^E(x)$ for $x$ without any 0-bit, otherwise
\begin{equation}
\mu^E(x) = CP^E(x) - CP^E(x').
\end{equation}
\end{corollary}

\subsection{Universal CEM vs EOM with Random Input}

Corollary \ref{muE<mu0} and Lemma \ref{mu0<muE}
below imply that $\mu^E$ and  $\mu_0$
are essentially the same thing: randomly selecting the inputs 
of a universal EOM
yields output prefixes whose probabilities are determined
by the universal CEM.

\begin{corollary}
\label{muE<mu0}
Let $\mu_0$ denote the universal CEM of Theorem \ref{universalCEM}.
For $x \in B^*$, 
\[
\mu^E(x) = O(\mu_0(x)). 
\]
\end{corollary}

\begin{lemma}
\label{mu0<muE}
For $x \in B^*$, 
\[
\mu_0(x)= O(\mu^E(x)).
\]
\end{lemma}
{\bf Proof.}
In the enumeration of EOMs in the proof of Theorem \ref{universalCEM},
let $EOM_0$ be an EOM representing $\mu_0$. 
We build an EOM $T$ such that $\mu_T(x) = \mu_0(x)$.
The rest follows from the Invariance Theorem
(compare Def. \ref{invariance}).

$T$ applies $EOM_0$ to all $x \in B^*$ in dovetail fashion,
and simultaneously
simply reads randomly selected input bits forever. 
At a given time, let string variable $z$ denote $T$'s input
string read so far.
Starting at the right end of the unit interval $[0,1)$,
as the $V\bar{\mu}_0(x)$ are being updated by the algorithm of 
Theorem \ref{universalCEM}, $T$ 
keeps updating a chain of finitely many, variable, disjoint,
consecutive, adjacent, half-open intervals 
$VI(x)$
of size $V\bar{\mu}_0(x)$ in alphabetic order on $x$,
such that $VI(y)$ is to the right of $VI(x)$ if $y \succ x$.  
After every variable update and each increase of 
$z$, $T$ replaces its output by the
$x$ of the $VI(x)$ with $0.z \in VI(x)$. 
Since neither $z$ nor the $VC\mu_0(x)$ 
in the algorithm of Theorem \ref{universalCEM}
can decrease (that is, all interval boundaries can only shift left),
$T$'s output cannot either, and therefore is indeed EOM-computable. 
Obviously the following holds:
\[
C\mu P_T(x) = 
CP_T(x) = 
C\mu_0(x)
\]
and 
\[
\mu P_T(x) = \sum_{z \in B^{\sharp}} P_T(xz) = \mu_0(x).
\]
$\Box$
\vspace{0.3cm}

\section{Probability vs Descriptive Complexity}
\label{coding}

The size of some computable object's minimal description is closely
related to the object's probability.  For instance,
Levin \cite{Levin:74} 
proved the remarkable {\em Coding Theorem}
for his universal discrete enumerable semimeasure 
$m$ based on halting programs (see Def. \ref{m}); compare independent
work by Chaitin \cite{Chaitin:75} 
who also gives credit to N. Pippenger:  

\begin{theorem}[Coding Theorem]
\label{codingm}
\begin{equation}
 For~x \in B^*,~
-log~m(x) \leq  K(x) \leq -log~m(x) + O(1)  
\end{equation}
\end{theorem}
In this special case, the contributions of the
shortest programs dominate the probabilities of objects
computable in the traditional sense.
As shown by G\'{a}cs \cite{Gacs:83} 
for the case of MTMs, however,
contrary to Levin's \cite{Levin:73a} conjecture,
$\mu^M(x) \neq O(2^{-Km^M(x)}); $ 
but a slightly worse bound does hold:
\begin{theorem}
\begin{equation}
K\mu^M(x) - 1 \leq Km^M(x) \leq K\mu^M(x) + Km^M(K\mu^M(x)) + O(1). 
\end{equation}
\label{KmuM}
\end{theorem}
The term $-1$ on the left-hand side stems from the definition of $lg(x) \leq log(x)$.
We will now consider the case of 
probability distributions that dominate $m$,
and semimeasures that dominate $\mu^M$,
starting with the case of enumerable objects.

\subsection{Theorems for EOMs and GTMs}

\begin{theorem}
\label{K-KP}
For $x \in B^{\sharp}$ with $P^E(x) > 0$,
\begin{equation}
KP^E(x) - 1 \leq  K^E(x) \leq  KP^E(x) + K^E(KP^E(x)) + O(1). 
\end{equation}
\end{theorem}
Using $K^E(y) \leq log~y + 2log~log~y + O(1)$ for $y$ interpreted
as an integer --- compare 
Def. \ref{encoding}  
--- this yields
\begin{equation}
2^{-K^E(x)} < P^E(x) \leq O(2^{-K^E(x)})(K^E(x))^2.
\end{equation}
That is, objects that are hard to describe 
(in the sense that they have only long enumerating descriptions)
have low probability. 
\vspace{0.3cm}

\noindent {\bf Proof.} 
The left-hand inequality follows by definition.
To show the right-hand side, one can build an EOM $T$ that computes $x \in
B^{\sharp}$ using not more than $KP^E(x) + K_T(KP^E(x)) + O(1)$ input
bits in a way inspired by Huffman-Coding \cite{Huffman:52}. 
The claim then follows from the Invariance Theorem.
The trick is to arrange $T$'s computation such that $T$'s output converges
yet never needs to decrease lexicographically.
$T$ works as follows: 

\begin{quote}
{\bf (A)} 
Emulate $U^E$
to construct a real enumerable number $0.s$ encoded as
a self-delimiting input program $r$, simultaneously 
run all (possibly
forever running) programs on $U^E$ dovetail style; whenever the output
of a prefix $q$ of any running program starts with some $x \in B^*$ for
the first time, set variable $V(x) := V(x) + 2^{-l(q)}$ (if no program
has ever created output starting with $x$ then first create $V(x)$
initialized by 0); whenever the output of some extension $q'$ of $q$
(obtained by possibly reading additional input bits: $q'=q$ if none are
read) lexicographically increases such that it does not equal $x$ any
more, set $V(x) := V(x) - 2^{-l(q')}$.

{\bf (B)} 
Simultaneously, starting at the right end of the unit interval $[0,1)$,
as the $V(x)$ are being updated, keep updating a chain of disjoint,
consecutive, adjacent, half-open (at the right end) intervals $IV(x) =
[LV(x), RV(x))$ of size $V(x) = RV(x) - LV(x)$ in alphabetic order on $x$,
such that the right end of the $IV(x)$ of the largest $x$ coincides with
the right end of $[0,1)$, and $IV(y)$ 
is to the right of $IV(x)$ if $y \succ x$.  After every
variable update and each change of $s$, replace the output of $T$ by the
$x$ of the $IV(x)$ with $0.s \in IV(x)$. 

\end{quote}
This will never violate the EOM
constraints: the enumerable $s$ cannot shrink, and since EOM outputs
cannot decrease lexicographically, the interval boundaries $RV(x)$ and
$LV(x)$ cannot grow (their negations are enumerable, compare Lemma 
\ref{CPEenumerable}), hence $T$'s output cannot decrease.

For $x \in B^*$ the $IV(x)$ converge towards an interval $I(x)$ of
size $P^E(x)$.  For $x \in B^{\infty}$ with $P^E(x) > 0$, we have: for any
$\epsilon >0$ there is a time $t_0$ such that for all time steps $t>t_0$
in $T$'s computation, an interval $I_{\epsilon}(x)$ of size $P^E(x) -
\epsilon$ will be completely covered by certain $IV(y)$ satisfying $x
\succ y$ and $0.x - 0.y < \epsilon$.  So for $\epsilon \rightarrow 0$ the
$I_{\epsilon}(x)$ also converge towards an interval $I(x)$ of size $P^E(x)$.
Hence $T$ will output larger and larger $y$ approximating $x$ from below,
provided $0.s \in I(x)$.

Since any interval of size $c$ within $[0,1)$ contains a number $0.z$
with $l(z)$ = $-lg~c$, in both cases there is a number $0.s$ (encodable
by some $r$ satisfying $r \leq l(s) + K_T(l(s)) + O(1))$) with $l(s)$ =
$-lg P^E(x) + O(1)$, such that $T(r) \leadsto x$, and therefore $K_T(x)
\leq l(s) + K_T(l(s)) + O(1)$.  
$\Box$
\vspace{0.5cm}

\noindent
Less symmetric statements can also be derived in very similar fashion:

\begin{theorem}
\label{KG-KPT}
Let TM $T$ induce {\em approximable} $CP_T(x)$
for all $x \in B^*$
(compare Defs. \ref{defCP} and \ref{continua}; an EOM would be a special case). 
Then for $x \in B^{\sharp}$, $P_T(x) > 0$:
\begin{equation}
K^G(x) \leq  KP_T(x) + K^G(KP_T(x)) + O(1). 
\label{KPT}
\end{equation}
\end{theorem}

\noindent {\bf Proof.} Modify the proof of Theorem \ref{K-KP} for approximable
as opposed to enumerable interval boundaries and approximable $0.s$.
$\Box$
\vspace{0.5cm}

\noindent
A similar proof, but without the complication for the case $x \in
B^{\infty}$, yields:

\begin{theorem}
\label{KmG-Kmu}
Let $\mu$ denote an approximable  semimeasure on $x \in B^*$; that is,
$\mu(x)$ is describable. Then
for $\mu(x) > 0$:
\begin{equation}
Km^G(x) \leq  K\mu(x) + Km^G(K\mu(x)) + O(1);
\end{equation}
\begin{equation}
K^G(x) \leq  K\bar{\mu}(x) + K^G(K\bar{\mu}(x)) + O(1).
\end{equation}
\end{theorem}
As a consequence, 
\begin{equation}
\frac{\mu(x)} 
     {K\mu(x)log^2K\mu(x)} 
\leq O(2^{-Km^G(x)}); ~~
\frac{\bar{\mu}(x)} 
     {K\bar{\mu}(x)log^2K\bar{\mu}(x)} 
\leq O(2^{-K^G(x)}).
\label{muKmuKmG}
\end{equation}

\noindent {\bf Proof.} 
Initialize variables $V_{\lambda} := 1$ and $IV_{\lambda} := [0,1)$.
Dovetailing over all $x \succ \lambda$,
approximate the GTM-computable $\bar{\mu}(x) = \mu(x)-\mu(x0) -\mu(x1)$ in
variables $V_x$ initialized by zero, and create a chain of adjacent
intervals $IV_x$ analogously to the proof of Theorem \ref{K-KP}.

The $IV_x$ converge against intervals $I_x$ of size $\bar{\mu}(x)$.  
Hence $x$
is GTM-encodable by any program $r$ producing an output $s$ with 
$0.s \in I_x$: 
after every update, replace the GTM's output by the $x$ of
the $IV_x$ with $0.s \in IV_x$. 
Similarly, if $0.s$
is in the union of adjacent intervals $I_y$ of strings $y$ starting with $x$,
then the GTM's output
will converge towards some string starting with $x$.
The rest follows in a way similar
to the one described in the 
final paragraph of the proof of Theorem \ref{K-KP}. $\Box$
\vspace{0.5cm}

%Compare Levin's "Strong Church's Thesis" (personal communication, Dec
%2000) which says that any experimental sequence must be independent of
%(i.e., have small mutual information with) any sequence with a short
%mathematical description.

\noindent
Using the basic ideas in the proofs of Theorem \ref{K-KP} and \ref{KmG-Kmu} 
in conjunction with Corollary \ref{muE<mu0} and Lemma \ref{mu0<muE}, 
one can also obtain statements such as:

\begin{theorem}
\label{Kmu-KmE}
Let $\mu_0$ denote the universal CEM from Theorem \ref{universalCEM}.
For $x \in B^*$,
\begin{equation}
K\mu_0(x) - O(1) \leq  Km^E(x) \leq  K\mu_0(x) + Km^E(K\mu_0(x)) + O(1) 
\label{KmE}
\end{equation}
\end{theorem}
While $P^E$ dominates $P^M$ and $P^G$ dominates $P^E$, the reverse
statements are not true. In fact, given the results from Sections
\ref{power} and \ref{coding}, one can now make claims such as
the following ones:

\begin{corollary}
\label{PG>PE}
The following functions are unbounded:
\[
\frac{\mu^E(x)}{\mu^M(x)}; ~~
\frac{P^E(x)}{P^M(x)}; ~~
\frac{P^G(x)}{P^E(x)}. 
\]
\end{corollary}

\noindent {\bf Proof.} 
For the cases $\mu^E$ and $P^E$, apply Theorems \ref{KmuM}, \ref{Kmu-KmE} and
the unboundedness of (\ref{KE-KM}).
For the case $P^G$, apply Theorems \ref{z} and \ref{K-KP}.
 
\subsection{Tighter Bounds?}

Is it possible to get rid of the small correction terms such as
$K^E(KP^E(x)) \leq O(log(-log P^E(x))$ in Theorem \ref{K-KP}? 
Note that the construction in the proof shows that $K^E(x)$ is actually
bounded by $K^E(s)$, the complexity of the enumerable number $0.s \in
I(x)$ with minimal $K_T(s)$.  The facts 
$\sum_x P^M(x) = 1$, $\sum_x P^E(x) = 1$, $\sum_x P^G(x) < 1$,
as well as
intuition and wishful thinking
inspired by Shannon-Fano Theorem \cite{Shannon:48} and Coding
Theorem \ref{codingm} suggest there might indeed be tighter bounds:

\vspace{0.5cm}

\begin{conjecture}
\label{KM?KP}
For $x \in B^{\sharp}$ with $P^M(x) > 0$: $K^M(x) \leq KP^M(x) + O(1)$.
\end{conjecture}

\begin{conjecture}
\label{KE?KP}
For $x \in B^{\sharp}$ with $P^E(x) > 0$: $K^E(x) \leq KP^E(x) + O(1)$.
\end{conjecture}

\begin{conjecture}
\label{KG?KP}
For $x \in B^{\sharp}$ with $P^G(x) > 0$: $K^G(x) \leq KP^G(x) + O(1)$.
\end{conjecture}
The work of G\'{a}cs has already shown, however, that analogue conjectures for 
semimeasures such as $\mu^M$ 
(as opposed to distributions)
are false \cite{Gacs:83}. 

\subsection{Between EOMs and GTMs?}
 
The dominance of $P^G$ over $P^E$ comes at
the expense of occasionally ``unreasonable,'' nonconverging outputs.
Are there classes of always converging TMs more expressive than EOMs?
Consider a TM called a
PEOM whose inputs are pairs of finite bitstrings $x,y \in B^*$ (code
them using $2log~l(x) + 2log~l(y)  + l(xy) + O(1)$ bits).  The 
PEOM uses dovetailing to run
all self-delimiting programs on the $y$-th EOM of an enumeration of all
EOMs, to approximate the probability $PEOM(y,x)$ (again encoded as a string)
that the EOM's output
starts with $x$.  $PEOM(y,x)$ is approximable (we may 
apply Theorem \ref{KmG-Kmu}) but
not necessarily enumerable.  On the other hand, it is easy to see that
PEOMs can compute all enumerable strings describable on EOMs. In this
sense PEOMs are more expressive than EOMs, yet never diverge like GTMs.
EOMs can encode some enumerable strings slightly more compactly, however,
due to the PEOM's possibly unnecessarily bit-consuming input encoding.
An interesting topic of future research may be to establish a
partially ordered expressiveness hierarchy among classes of always
converging TMs, and to characterize its top, if there is one, which we doubt.
Candidates to consider may include TMs 
that approximate certain recursive or enumerable functions 
of enumerable strings.

\section{Temporal Complexity}
\label{temporalcomp}

So far we have completely ignored the time necessary to compute objects
from programs. 
In fact, the objects that are highly probable
according to $P^G$ and $P^E$ and $\mu^E$ 
introduced in the previous sections yet
quite improbable according to less dominant priors studied earlier (such
as $\mu^M$ and recursive 
priors \cite{Zvonkin:70,Levin:74,Solomonoff:78,Gacs:83,LiVitanyi:97}) 
are precisely those whose computation
requires immense time. For instance, the
time needed to compute 
the describable, even enumerable $\Omega_n$ 
grows faster than any recursive function of $n$, as shown by Chaitin
\cite{Chaitin:87}. 
Analogue statements hold for the $z$ of Theorem \ref{K-KG}.
Similarly, many of the semimeasures discussed above 
are approximable, but the approximation process 
is excessively time-consuming.

Now we will study the opposite extreme, 
namely, priors with a bias towards
the fastest way of producing certain outputs.
Without loss of generality, we will focus on computations
on a universal MTM.  For simplicity let us extend the
binary alphabet such that it contains an additional 
output symbol ``blank.''

\subsection{Fast Computation of Finite and Infinite Strings}
\label{fastinfinite}

There are many ways of systematically enumerating all computable objects
or bitstrings.  All take infinite time. Some, however, compute individual
strings much faster than others. To see this, first consider the trivial
algorithm ``ALPHABET,'' which simply lists all bitstrings ordered by size
and separated by blanks
(compare Marchal's thesis \cite{Marchal:98} and Moravec's library of all
possible books \cite{Moravec:99}).  ALPHABET will eventually
create all initial finite segments of all strings.  For example, the
$n$th bit of the string ``11111111...'' will appear as part of ALPHABET's
$2^n$-th output string.  Note, however, that countably many steps
are {\em  not} sufficient to print any infinite string of countable size!

There are much faster ways though.  For instance, the algorithm used in
the previous paper on the computable universes \cite{Schmidhuber:97brauer}
sequentially computes all computable bitstrings by a particular form of
dovetailing.  Let $p^i$ denote the $i$-th possible program.  Program $p^1$
is run for one instruction every second step (to simplify things, if the
TM has a halt instruction and $p^1$ has halted we assume nothing is done
during this step --- the resulting loss of efficiency is  not significant
for what follows). Similarly, $p^2$ is run for one instruction every
second of the remaining steps, and so on.

Following Li and Vit\'{a}nyi \cite[p. 503 ff]{LiVitanyi:97}, 
let us call this popular dovetailer
``SIMPLE.''  It turns out that SIMPLE actually is the fastest in a certain
sense.  For instance, the $n$th bit of string ``11111111...'' now will
appear after at most $O(n)$ steps (as opposed to at least $O(n2^n)$ steps
for ALPHABET). Why?  Let $p^k$ be the fastest algorithm that outputs ``11111111...''.
Obviously $p^k$ computes the $n$-th bit within $O(n)$ instructions.
Now SIMPLE will execute one instruction of $p^k$ every $2^{-k}$ steps.
But $2^{-k}$ is a positive constant that does not depend on $n$.

Generally speaking, suppose $p^k$ is among the fastest finite algorithms
for string $x$ and computes $x_n$ 
within at most $O(f(n))$ instructions, for all $n$. 
Then $x$'s first $n$ symbols will appear after at most $O(f(n))$ steps of SIMPLE.
In this sense SIMPLE essentially computes each string as quickly as
its fastest algorithm, although it is in fact computing all computable
strings simultaneously. This may seem counterintuitive.

\subsection{FAST: The Most Efficient Way of Computing Everything}
\label{FAST}

Subsection \ref{fastinfinite} 
focused on SIMPLE ``steps'' allocated for
instructions of single string-generating algorithms. Note that each such
step may require numerous ``micro-steps'' for the computational overhead
introduced by the need for organizing internal storage.  For example,
quickly growing space requirements for storing all strings may
force a dovetailing TM to frequently shift its writing and scanning
heads across large sections of its internal tapes.  This may consume
more time than necessary.

To overcome potential slow-downs of this kind, and to optimize the
TM-specific ``constant factor,'' we will slightly modify an 
optimal search algorithm
called {\em ``Levin search''} \cite{Levin:73,Levin:84,Adleman:79,LiVitanyi:97}
(see \cite{Schmidhuber:97nn,Wiering:96levin,Schmidhuber:97bias} 
for the first practical applications we are aware of).
Essentially, we will strip Levin search of its search aspects and apply it
to possibly infinite objects.  This leads to the most efficient (up
to a constant factor depending on the TM) algorithm for computing all
computable bitstrings.

\begin{quote}
 {\bf FAST Algorithm }:
For $i = 1, 2, \ldots $  perform PHASE $i$:

\begin{quote}
PHASE $i$: Execute $2^{i - l(p)}$ instructions of
all program prefixes $p$ satisfying
$l(p) \leq i$, and sequentially write the outputs on adjacent
sections of the output tape, separated by blanks.
\end{quote}

\end{quote}
Following Levin \cite{Levin:73},
within $2^{k+1}$ TM steps, each of order $O(1)$ ``micro-steps''
(no excessive computational overhead due to storage allocation etc.),
{\bf FAST} will generate all prefixes $x_n$
satisfying $Kt(x_n) \leq  k$, where $x_n$'s Levin complexity $Kt(x_n)$ is
defined as
\[
Kt(x_n) = \min_q\{ l(q) + log~t(q,x_n)\},
\]
where program prefix $q$ computes $x_n$ in $t(q,x_n)$ time steps.
The computational complexity of the algorithm
is not essentially affected by the fact
that PHASE $i = 2, 3, \ldots $, 
repeats the computation of PHASE $i-1$ which for large
$i$ is approximately half as short (ignoring nonessential
speed-ups due to halting programs if there are any).

One difference between SIMPLE and {\bf FAST} is that SIMPLE may
allocate steps to algorithms with a short description less frequently
than {\bf FAST}.  Suppose no finite algorithm computes $x$ faster than
$p^k$ which needs at most $f(n)$ instructions for $x_n$, for all $n$.
While SIMPLE needs $2^{k+1} f(n)$ steps to compute $x_n$, following
Levin \cite{Levin:73} it can be shown that {\bf FAST} requires at most
$2^{K(p^k)+1}f(n)$ steps --- compare \cite[p. 504 ff]{LiVitanyi:97}. 
That is, SIMPLE and {\bf FAST} share the same
order of time complexity (ignoring SIMPLE's ``micro-steps'' for storage
organization), but {\bf FAST}'s constant factor tends to be better.

Note that an observer $A$ evolving in one of the universes 
computed by {\bf FAST} might decide to build a machine that simulates all
possible computable universes using {\bf FAST}, and so on, recursively.
Interestingly, this will not necessarily cause a dramatic exponential 
slowdown: if the $n$-th discrete time step of $A$'s universe 
(compare Example \ref{universe}) is computable within $O(n)$ time then
$A$'s simulations can be as fast as the ``original'' simulation,
save for a constant factor. In this sense a ``Great Programmer''
\cite{Schmidhuber:97brauer} who writes a program that runs all possible
universes would not be superior to certain nested Great Programmers 
inhabiting his universes.

To summarize: the effort required for computing all computable objects
simultaneously does not prevent {\bf FAST} from computing each object
essentially as quickly as its fastest algorithm.  No other dovetailer
can have a better order of computational complexity.  This suggests a
notion of describability that is much more restricted yet perhaps much
more natural than the one used in the earlier sections on description
size-based  complexity. 

\subsection{Speed-Based Characterization of the Describable}
\label{characterization}

The introduction mentioned that some sets seem describable in a certain
sense while most of their elements are not. Although the dyadic
expansions of most real numbers are not individually describable, the
short algorithm ALPHABET from Section \ref{fastinfinite} will compute all
their finite prefixes.  However, ALPHABET is unable to
print {\em any} infinite string using only countable time and storage.
Rejection of the notion of uncountable storage and time
steps leads to a speed-based definition of describability.

\begin{definition}[``S-describable'' Objects] 
\label{quickly}
Some $x \in B^{\sharp}$ is 
{\em S-describable} (``S'' for ``Speed'') if it has a
finite algorithm that outputs $x$ using countable time and space.
\end{definition}

\begin{lemma} 
\label{countable} 
With countable time and space requirements, {\bf FAST} 
computes all S-describable strings. 
\end{lemma}
To see this, 
recall that {\bf FAST} will output any S-describable string
as fast as its fastest algorithm, save for a constant factor.
Those $x$ with polynomial time bounds on the computation of $x_n$ (e.g.,
$O(n^{37})$) are S-describable, but most $x \in B^{\sharp}$ are not,
as obvious from Cantor's insight \cite{Cantor:1874}.

The prefixes $x_n$ of all $x \in B^{\sharp}$, even of those that are not
S-describable, are computed within at most $O(n2^n)$ steps, at least as
quickly as by ALPHABET. The latter, however, never is faster than that,
while {\bf FAST} often is.  Now consider infinite strings $x$ whose
fastest {\em individual} finite program needs even more than $O(n2^n)$
time steps to output $x_n$ and nothing but $x_n$, such as Chaitin's
$\Omega$ (or the even worse $z$ from Theorem \ref{z}) --- recall that the
time for computing $\Omega_n$ grows faster than any recursive function
of $n$ \cite{Chaitin:87}.  We observe that this result is irrelevant for
{\bf FAST} which will output $\Omega_n$ within $O(n2^n)$ steps, but only
because it also outputs many other strings besides $\Omega_n$ --- there
is still no fast way of identifying $\Omega_n$ among all the outputs.
$\Omega$ is not S-describable because it is not generated any more quickly
than uncountably many other infinite and incompressible strings, which
are not S-describable either.

\subsection{Enumerable Priors vs FAST }

The {\bf FAST} algorithm gives rise to a natural prior 
measure on the computable objects which is much less
dominant than $\mu^M$, $\mu^E$ and $\mu^G$. This prior will be
introduced in Section \ref{speedprior} below.
Here we first motivate it by evaluating drawbacks
of the traditional, well-studied, enumerable prior $\mu^M$ 
\cite{Solomonoff:64,Levin:74,Solomonoff:78,Gacs:83,LiVitanyi:97} 
in the context of {\bf FAST}.

\begin{definition}[$p \to x, p \to_i x $] 
\label{p->x} 
Given program prefix $p$, write $p \to x$ if
our MTM reads $p$ and computes output starting with $x \in B^*$,
while no prefix of $p$ consisting of less than $l(p)$ bits
outputs $x$. Write
$p \to_i x$ if
$p \to x$ in PHASE $i$ of {\bf FAST}.
\end{definition} 
We observe that
\begin{equation}
\mu^M(x) = lim_{i \to \infty} \sum_{p \to_i x} 2^{-l(p)},
\end{equation}
but there is no recursive function $i(x)$ such that
\begin{equation}
\mu^M(x) = \sum_{p \to_{i(x)} x} 2^{-l(p)},
\end{equation}
otherwise $\mu^M(x)$ would be recursive.
Therefore we might argue that the use of prior
$\mu^M$ is essentially equivalent to using a probabilistic version
of {\bf FAST} which 
randomly selects a phase according to a distribution assigning
zero probability to any phase with recursively computable number.  
Since the time
and space consumed by PHASE $i$ is at least $O(2^i)$, we are 
approaching uncountable resources as $i$ goes to infinity.
From any reasonable computational perspective, however, the probability
of a phase consuming more than countable resources clearly should be zero.
This motivates the next subsection.

\subsection{Speed Prior $S$ and Algorithm GUESS }
\label{speedprior}

A resource-oriented point of view 
suggests the following postulate.

\begin{postulate}
\label{postulate}
The cumulative prior probability measure of all $x$ 
incomputable within time $t$
by the most efficient way of computing everything
should be inversely proportional to $t$.
\end{postulate}
Since  the most efficient way of computing all $x$ is embodied by {\bf
FAST}, and since each phase of {\bf FAST} consumes roughly twice the
time and space resources of the previous phase, the cumulative prior
probability of each finite phase should be roughly half the one of
the previous phase; zero probability should be assigned to infinitely
resource-consuming phases.  Postulate \ref{postulate} therefore suggests the
following definition.

\begin{definition}[Speed Prior $S$]
\label{S}
Define the speed prior $S$ on $B^*$ as
\[
S(x) :=  \sum_{i=1}^{\infty} 2^{-i} S_i(x);~~ where~
S_i(\lambda) = 1;~
S_i(x) = \sum_{p \to_i x} 2^{-l(p)} ~for~x \succ \lambda.
\]
\end{definition}
We observe that $S(x)$ is indeed a semimeasure (compare Def. \ref{mu}):
\[
S(x0) + S(x1) + \bar{S}(x) = S(x); ~~where~\bar{S}(x)  \geq 0.
\]
Since $x \in B^*$ is first computed in PHASE $Kt(x)$ within 
$2^{Kt(x)+1}$ steps, we may rewrite:
\begin{equation}
S(x) = 2^{-Kt(x)} \sum_{i=1}^{\infty} 2^{-i} S_{Kt(x)+i-1}(x)
\leq 2^{-Kt(x)} 
\end{equation}
$S$ can be implemented by the following 
probabilistic algorithm for a universal MTM.

\begin{quote}
Algorithm {\bf GUESS}:
\begin{quote}

{\bf 1.} 
Toss an unbiased coin until heads is up;
let $i$ denote the number of required trials;
set $t:=2^i$. 

{\bf 2.} 
If the number of steps executed so far exceeds $t$ then exit.
Execute one step; if it is a request for an input bit, 
toss the coin to determine the bit, and set $t:=t/2$. 

{\bf 3.} Go to {\bf 2.}

\end{quote}
\end{quote}
In the spirit of {\bf FAST},
algorithm {\bf GUESS} makes twice the computation time half as likely,
and splits remaining time in half whenever a new bit is requested, 
to assign equal runtime to the two resulting sets of possible
program continuations.
Note that the expected runtime of {\bf GUESS} is unbounded since
$\sum_i 2^{-i} 2^{i}$ does not converge. Expected runtime is countable,
however, and expected space is of the order of expected time, 
due to numerous short algorithms producing a constant
number of output bits per constant time interval.

Assuming our universe is sampled according to GUESS implemented on some
machine, note that the true distribution is not essentially different
from the estimated one based on our own, possibly different  machine.

\subsection{Speed Prior-Based Inductive Inference}
\label{Sbasedinference}

Given $S$, as we observe an initial segment $x \in B^*$ of some string,
which is the most likely continuation?
Consider $x$'s finite continuations $xy, y \in B^*$.
According to Bayes (compare Equation (\ref{bayes2})),
\begin{equation}
S(xy \mid x) = \frac{S(x \mid xy) S(xy)} {S(x)}
= \frac{S(xy)} {S(x)},
\label{Sbayes}
\end{equation}
where $S(z^2 \mid z^1)$ is the measure of $z^2$, given $z^1$. 
Having observed $x$ we will predict those $y$ that maximize 
$S(xy \mid x)$.  Which are those? In what follows,
we will confirm the intuition that 
for $n \rightarrow \infty$ the
only probable continuations of $x_{n}$  are those with fast
programs.  The sheer number of ``slowly''
computable strings cannot balance the speed advantage
of ``more quickly'' computable strings with equal beginnings.

\begin{definition}[$p \stackrel{< k}{\longrightarrow}_i x$ etc.]
Write $p \stackrel{< k}{\longrightarrow} x$ if
finite program $p$ ($p \to x$) computes  $x$ within 
less than $k$ steps, and 
$p \stackrel{< k}{\longrightarrow_i} x$ if it does so within PHASE $i$ of {\bf FAST}.
Similarly for
$p \stackrel{\leq k}{\longrightarrow} x$ 
and $p \stackrel{\leq k}{\longrightarrow_i} x$ 
(at most $k$ steps), 
$p \stackrel{= k}{\longrightarrow} x$,
(exactly $k$ steps), 
$p \stackrel{\geq k}{\longrightarrow} x$,
(at least $k$ steps), 
$p \stackrel{> k}{\longrightarrow} x$  (more  than $k$ steps). 
\end{definition}

\begin{theorem}
\label{probablyfast}
Suppose $x \in B^{\infty}$ is S-describable, and
$p^x \in B^*$ outputs $x_{n}$ within at most $f(n)$ steps for all $n$, 
and
$
g(n) > O(f(n)).
$
Then 
\[
Q(x,g,f):=
lim_{n \to \infty} 
\frac{ \sum_{i=1}^{\infty} 2^{-i} \sum_{p \stackrel{\geq g(n)}{\longrightarrow_i} x_n} 2^{-l(p)}}
     { \sum_{i=1}^{\infty} 2^{-i} \sum_{p \stackrel{\leq f(n)}{\longrightarrow_i} x_n} 2^{-l(p)}}
= 0.
\]
\end{theorem}

\noindent {\bf Proof.} Since no program that requires at least $g(n)$ steps for producing
$x_n$ can compute $x_n$ in a phase with number $ < log~g(n)$, we have
\[
Q(x,g,f) \leq
lim_{n \to \infty} 
\frac
{\sum_{i=1}^{\infty} 2^{-log~g(n)-i} 
\sum_{p \stackrel{\geq g(n)}{\longrightarrow}_{(i+log~g(n))} x_n}2^{-l(p)}}
{\sum_{i=1}^{\infty} 2^{-log~f(n)-i} 
\sum_{p \stackrel{= f(n)}{\longrightarrow_i} x_n}2^{-l(p)}} 
\leq
\]
\[
lim_{n \to \infty} 
\frac{ f(n) \sum_{p \to x_n}   2^{-l(p)}  }
     { g(n) \sum_{p \stackrel{= f(n)}{\longrightarrow} x_n}   2^{-l(p)}  } \leq
lim_{n \to \infty} 
\frac{ f(n) }
     { g(n) } 
\frac{ 1 }
     { 2^{-l(p^x)}  } = 0.
\]
Here we have used the Kraft inequality \cite{Kraft:49} 
to obtain a rough upper bound for the enumerator: 
when no $p$ is prefix of another one, then $\sum_{p} 2^{-l(p)} \leq 1.$
$\Box$

\vspace{0.3cm}
\noindent
Hence, if we know a rather fast
finite program $p^x$ for $x$, then Theorem \ref{probablyfast} allows for predicting: 
if we observe some $x_n$ ($n$ sufficiently large) then
it is very unlikely that it was produced by an $x$-computing algorithm
much slower than $p^x$.  

Among the fastest algorithms for $x$ is {\bf FAST} itself, which is 
at least as fast as $p^x$, save for a constant factor. It outputs
$x_n$ after $O(2^{Kt(x_n)})$ steps. Therefore Theorem \ref{probablyfast} tells us:

\begin{corollary}
\label{probablyKt}
Let $x \in B^{\infty}$ be S-describable.
For $n \to \infty$, with probability 1 the 
continuation of $x_n$
is computable within $O(2^{Kt(x_n)})$ steps.
\end{corollary}
Given observation $x$ with $l(x) \to \infty$, we
predict a continuation $y$ with minimal $Kt(xy)$.

\begin{example}
\label{duration}
{\em 
Consider 
Example \ref{whereamI}  and Equation (\ref{bayes1}).
According to the weak anthropic principle,
the conditional probability of a particular observer finding herself 
in one of the universes compatible with her existence equals 1.  
Given $S$, we predict a universe with minimal $Kt$.
Short futures are more likely than long ones: the probability
that the universe's history so far will extend beyond the one computable
in the current phase of {\bf FAST} (that is, it will be prolongated into
the next phase) is at most 50 \%. Infinite futures have measure zero.
}
\end{example}

\subsection{Practical Applications of Algorithm GUESS}

Algorithm {\bf GUESS} is almost identical to a probabilistic search
algorithm used in previous work on applied inductive inference
\cite{Schmidhuber:95kol,Schmidhuber:97nn}. The 
programs generated by the previous algorithm, however,
were not bitstrings but written in an assembler-like language;
their runtimes had an upper bound, and the
program outputs were evaluated as to whether they represented solutions
to externally given tasks.

Using a small set of exemplary training examples, the system 
discovered the weight matrix of an artificial neural network whose task
was to map input data to appropriate target classifications.  The network's
generalization capability was then
tested on a much larger unseen test set. On several toy problems it generalized
extremely well in a way unmatchable by traditional neural network
learning algorithms.

The previous papers, however, did not explicitly establish
the above-mentioned relation between ``optimal'' resource 
bias and {\bf GUESS}.

\section{Consequences for Physics}
\label{consequences}

As obvious from equations (\ref{bayes1}) and (\ref{bayes2}), 
some observer's future depends
on the prior from which his/her universe is sampled.
More or less general
notions of TM-based describability put forward above lead to
more or less dominant priors such as 
$P^G$ on formally describable universes, 
$P^E$ and $\mu^E$ on enumerable universes, 
$P^M$ and $\mu^M$ and recursive priors on monotonically computable universes,
$S$ on S-describable universes. 
We will now comment on the plausibility of each, and discuss some consequences.  
Prior $S$, the arguably most plausible and natural
one, provokes specific predictions concerning our future.
 For a start, however, we will briefly review Solomonoff's traditional theory of
inductive inference based on {\em recursive} priors. 

\subsection{Plausibility of Recursive Priors}

The first number is 2, the second is 4, the third is 6, the fourth
is 8.  What is the fifth?  The correct answer is ``250,'' because
the $n$th number is $n^5 - 5n^4 -15n^3 + 125n^2 -224n + 120$.
In certain IQ tests, however,
the answer ``250'' will not yield maximal score, because it does
not seem to be the ``simplest'' answer consistent with the data
(compare \cite{Schmidhuber:97nn}).  
And physicists and others favor ``simple'' explanations of observations.

Roughly fourty years ago Solomonoff set out to provide a theoretical
justification of this quest for simplicity \cite{Solomonoff:64}.
He and others have made substantial progress over the past decades.
In particular, technical problems of Solomonoff's original approach were
partly overcome by Levin \cite{Levin:74} 
who introduced self-delimiting
programs, $m$ and $\mu^M$ mentioned above, as well as several
theorems relating probabilities to complexities --- see also 
Chaitin's and G\'{a}cs' 
independent papers on prefix complexity and $m$ 
\cite{Gacs:74,Chaitin:75}.
Solomonoff's work on inductive inference
helped to inspire less general yet practically more
feasible principles of 
minimum description length \cite{Wallace:68,Rissanen:86,Hochreiter:97nc1}
as well as time-bounded restrictions of
Kolmogorov complexity, e.g.,  \cite{Hartmanis:83,Allender:92,Watanabe:92,LiVitanyi:97},
as well as the concept of ``logical depth'' of $x$, the runtime of the shortest 
program of $x$ \cite{Bennett:88}.

Equation  (\ref{bayes2}) makes predictions of the entire future, given
the past.  This seems to be the most general approach.  Solomonoff 
\cite{Solomonoff:78} focuses just on the next
bit in a sequence.  Although this provokes surprisingly nontrivial
problems associated with translating the bitwise approach to alphabets
other than the binary one --- only recently Hutter managed to do this
\cite{Hutter:00e} --- it is sufficient for obtaining essential insights 
\cite{Solomonoff:78}.

Given an observed bitstring $x$, 
Solomonoff assumes the data are drawn according to
a recursive measure $\mu$; that is, there is a MTM program that 
reads $x \in B^*$ and computes $\mu(x)$ and halts. He
estimates the probability
of the next bit (assuming there will be one), using the fact
that the enumerable $\mu^M$ dominates the less general recursive measures:
\begin{equation}
K\mu^M(x) \leq -log \mu (x) + c_{\mu}, 
\end{equation}
where $c_{\mu}$ is a constant depending on $\mu$ but not on $x$.
Compare \cite[p. 282 ff]{LiVitanyi:97}.
Solomonoff showed that the $\mu^M$-probability of a particular
continuation converges
towards $\mu$ as the observation size goes to
infinity \cite{Solomonoff:78}.  
Hutter recently extended his results
by showing that
the number of prediction errors made by universal Solomonoff prediction 
is essentially bounded by the number of errors made by any other 
recursive prediction scheme, including the optimal scheme 
based on the true distribution $\mu$ \cite{Hutter:99}. 
Hutter also extended Solomonoff's passive universal induction framework
to the case of agents actively interacting with an unknown environment
\cite{Hutter:00f}.

A previous paper on computable universes
\cite[Section: {\em Are we Run by a Short Algorithm?}]{Schmidhuber:97brauer} 
applied the theory of inductive
inference to entire universe histories, and predicted that 
simple universes are more likely; that is, observers are 
likely to find themselves in a simple universe compatible
with their existence 
(compare {\em everything mailing list archive} \cite{Dai:98}, 
messages dated
21 Oct and 25 Oct 1999: 
{\em http://www.escribe.com/science/theory/m1284.html} and {\em m1312.html}).
There are two ways in which one could criticize this approach.
One suggests it is too general, the other suggests it is too restrictive.

{\bf 1. Recursive priors too general?}
$\mu^M(x)$ is not recursively computable, hence there is no general
practically feasible algorithm to generate optimal predictions.
This suggests to look at more 
restrictive priors, in particular, $S$, 
which will receive additional motivation further below.

{\bf 2. Recursive priors too restricted?} If we want to explain
the entire universe, then the assumption of a recursive
$P$ on the possible universes may even be insufficient.  In particular,
although our own universe seems to obey simple rules 
--- a discretized version of  
Schr\"{o}dinger's wave function could 
be implemented by a simple recursive algorithm --- the apparently noisy
fluctuations that we observe on top of the simple laws might be due to
a pseudorandom generator (PRG) subroutine whose output is describable,
even enumerable, but not recursive --- compare Example \ref{universe2}.

In particular, the fact that nonrecursive priors may not allow for
recursive bounds on the time necessary to compute initial histories of
some universe does not necessarily prohibit nonrecursive priors. Each
describable initial history may be potentially relevant as there
is an infinite computation during which it will be stable for 
all but finitely many steps.
This suggests to look at more general priors such as
$\mu^E, P^E, P^G$, which will be done next, before we come back to the
speed prior $S$.

\subsection{Plausibility of Cumulatively Enumerable Priors} 

The semimeasure $\mu^M$ 
used in the traditional theory of inductive inference 
is dominated by the nonenumerable yet approximable $\mu^E$
(Def. \ref{muT})
assigning approximable probabilities to initial segments of 
strings computable on EOMs.

As Chaitin points out \cite{Chaitin:87}, 
enumerable objects such as the halting probabilities of
TMs are already expressive enough to express anything
provable by finite proofs,
given a set of mathematical axioms. In particular, knowledge of
$\Omega^T_{n}$, the first $n$ bits of the halting probability of TM $T$,
conveys all information necessary to decide by a halting program whether
any given statement of an axiomatic system describable by fewer than $n-O(1)$ 
bits is provable or not within the system.

$\Omega^T_{n}$ is effectively random in the sense of Martin-L\"{o}f
\cite{Martin-Loef:66}.  Therefore it is generally undistinguishable
from noise by a recursive function of $n$, and thus very compact
in a certain sense --- in fact, all effectively random reals are
Omegas, as recently shown by Slaman \cite{Slaman:99} building on
work by Solovay \cite{Solovay:75}; see also
\cite{Calude:00,Solovay:00}.  One could still say, however, that
$\Omega$ {\em de}compresses mathematical truth at least enough to
make it retrievable by a halting program.  Assuming that this type
of mathematical truth contains everything relevant for a theory of
all reasonable universes, and assuming that the describable yet even
``more random''  patterns of Theorem \ref{z} are not necessary for such
a theory, we may indeed limit ourselves to the enumerable universes.

If Conjecture \ref{KE?KP} were true, then we would have $P^E(x) =
O(2^{-K^E(x)})$ (compare Equation (\ref{bayes1})), or $P^E(xy) =
O(2^{-K^E(xy)})$ (compare (\ref{bayes2})).  That is, the most likely
continuation $y$ would essentially be the one corresponding to the
shortest algorithm, and no cumulatively enumerable distribution could
assign higher probability than $O(2^{-K^E(xy)})$  to $xy$.  Maximizing
$P^E(xy)$ would be equivalent to minimizing $K^E(xy)$.

Since the upper bound given by Theorem \ref{K-KP} is not quite as sharp due to the
additional, at most logarithmic term, we cannot make quite as 
strong a statement. Still, Theorem \ref{K-KP} does tell us that $P^E(xy)$ goes
to zero with growing $K^E(xy)$ almost exponentially fast, 
and Theorem \ref{Kmu-KmE} says that $\mu^E(xy_k)$ ($k$ fix) goes
to zero with growing $Km^E(xy_k)$ almost exponentially fast. 

Hence, the relatively mild assumption that the probability distribution
from which our universe is drawn is cumulatively enumerable provides
a theoretical justification of the prediction that the most likely
continuations of our universes are computable by short EOM algorithms.
However, given $P^E$, 
Occam's razor (e.g., \cite{Blumer:87})
is only partially justified because
the {\em sum} of the probabilities of 
the most complex $xy$ does not vanish:
\[
lim_{n \rightarrow \infty} \sum_{xy \in B^{\sharp}: K^E(xy)>n}P^E(xy) > 0.
\]
To see this, compare 
Def. \ref{continua} and the subsequent
paragraph on program continua. 
There would be a nonvanishing
chance for an observer to end up in one of the maximally complex 
universes compatible with his existence, although only universes
with finite descriptions have nonvanishing individual probability.

We will conclude this subsection by addressing the issue of
falsifiability.  If $P^E$ or $\mu^E$ were responsible for the
pseudorandom aspects of our universe (compare Example \ref{universe2}),
then this might indeed be effectively undetectable in principle,
because some approximable and enumerable  patterns cannot be proven to
be nonrandom in recursively bounded time. Therefore the results above
may be of interest mainly from a philosophical point of view, not from a
practical one: yes, universes computable by short EOM algorithms are much 
more likely indeed, but even if we inhabit one then we may not be able to find its
short algorithm.

\subsection{Plausibility of Approximable Priors}

$\mu^E$ assigns low probability to G-describable strings such as the $z$ of
Theorem \ref{z}.  However, one might believe in the potential significance of such 
constructively describable patterns, 
e.g., by accepting their validity as possible pseudorandom perturbations
of a universe otherwise governed by a quickly computable
algorithm implementing simple physical laws --- compare Example \ref{universe2}.
Then one must also 
look at semimeasures dominating $\mu^E$, although
the falsifiability problem mentioned above holds for those as well.

The top of the TM dominance hierarchy is embodied by $G$ (Theorem \ref{z});
the top of our prior dominance hierarchy by $P^G$,
the top of the corresponding semimeasure dominance hierarchy by $\mu^G$.
If Conjecture \ref{KG?KP}
were true, then maximizing $P^G(xy)$ would be equivalent
to minimizing $K^G(xy)$.  Even then there would be a fundamental
problem besides lack of falsifiability: Neither $P^G$ nor $\mu^G$
are describable,  and not even a ``Great Programmer''
\cite{Schmidhuber:97brauer} could generally decide whether some GTM output
is going to converge (Theorem \ref{fluctuate}), or whether it actually represents a
``meaningless'' universe history that never stabilizes.

Thus, if one adopts the belief that nondescribable measures do not exist,
simply because there is no way of describing them, then one may discard
this option. 

This would suggest considering semimeasures less dominant than $\mu^G$, for
instance, one of the most dominant approximable $\mu$.  According to
Theorem \ref{KmG-Kmu} and inequality (\ref{muKmuKmG}), $\mu(xy)$ goes to
zero almost exponentially  fast with growing $Km^G(xy)$.  

As in the case of $\mu^E$, this may interest the philosophically inclined
more than the pragmatists: yes, any particular universe history without
short description necessarily is highly unlikely; much more likely are
those histories where our lives are deterministically computed by a short
algorithm, where the algorithmic entropy (compare \cite{Zurek:89b}) of
the universe does {\em not} increase over time, because a finite program
conveying a finite amount of information is responsible for everything,
and where concepts such as ``free will'' are just an illusion in a
certain sense.  Nevertheless, there may not be any effective way of
proving or falsifying this.

\subsection{Plausibility of Speed Prior $S$}
\label{Splausibility}

Starting with the traditional case of recursive priors, the subsections
above discussed more and more dominant priors as candidates for the
one from which our universe is sampled.  Now we will move towards the
other extreme: the less dominant prior $S$ which in a sense
is optimal with respect to temporal complexity.

So far, without much ado, we have used a terminology according to
which we ``draw a universe from a particular prior distribution.''
In the TM-based set-ups (see Def. \ref{continua}) this in principle
requires a ``binary oracle,'' a source of true randomness, to provide
the TM's inputs.  Any source of randomness, however, leaves us with an
unsatisfactory explanation of the universe, since random strings do not
have a compact explanation, by definition.  The obvious way around this,
already implicit in the definition of $\mu_T(x)$ (see Def.  \ref{muT}), is
the ``ensemble approach'' which runs {\em all} possible TM input programs and
sums over the lengths of those that compute strings starting with $x$.

Once we deal with ensemble approaches and explicit computations in
general, however, we are forced to accept their fundamental time
constraints.  As mentioned above, many of the shortest programs of
certain enumerable or describable strings compute their outputs more
slowly than any recursive upper bound could indicate.

If we do assume that time complexity of the computation should be
an issue, then why stop with the somewhat arbitrary restriction
of recursiveness, which just says that the time required to compute
something should be computable by a halting program?  Similarly, why
stop with the somewhat arbitrary restriction of polynomial time bounds
which are subject of much of the work in theoretical computer science?

If I were a ``Great Programmer'' \cite{Schmidhuber:97brauer} with
substantial computing resources, perhaps beyond those possible in our own
universe which apparently does not permit more than $10^{51}$ operations
per second and kilogram \cite{Bremermann:82,Lloyd:00}, yet constrained by the
fundamental limits of computability, I would opt for the fastest way of
simulating all universes, represented by algorithm {\bf FAST} 
(Section \ref{temporalcomp}).  Similarly,
if I were to search for some computable object with certain properties
discoverable by Levin's universal search algorithm (the ``mother'' of
{\bf FAST}), I would use the latter for its optimality properties.

Consider the observers evolving in the many different possible universes
computed by {\bf FAST} or as a by-product of Levin Search. Some of
them would be identical, at least for some time,
collecting identical experiences in universes with possibly equal
beginnings yet possibly different futures.  At a given time, the most
likely instances of a particular observer $A$ would essentially be
determined by the fastest way of computing $A$.

Observer $A$ might adopt the belief the Great Programmer was indeed smart
enough to implement the most efficient way of computing everything.
And given $A$'s very existence, $A$ can conclude that the Great
Programmer's resources are sufficient to compute at least one instance
of $A$.  What $A$ does not know, however, is the current phase of {\bf
FAST}, or whether the Great Programmer is interested in or aware of $A$,
or whether $A$ is just an accidental by-product of some Great Programmer's
search for something else, etc.

Here is where a resource-oriented bias comes in naturally.  It seems to
make sense for $A$ to assume 
that the Great Programmer is also bound by the limits of computability, 
that infinitely late phases of {\bf FAST}
consuming uncountable resources are infinitely unlikely, 
that any Great Programmer's {\em a priori} probability of 
investing computational resources into some search problem
tends to decrease with growing search costs,
and that the prior probability of anything whose computation 
requires more than $O(n)$ resources by the optimal method 
is indeed inversely proportional to $n$.  This immediately
leads to the speed prior $S$.

Believing in $S$, $A$ could use Theorem \ref{probablyfast} to predict
the future (or ``postdict'' unknown aspects of the past) 
by assigning highest probability to those S-describable
futures (or pasts) that are (a) consistent with $A$'s experiences and (b) are
computable by short and fast algorithms.  The appropriate simplicity
measure minimized by this resource-oriented version of Occam's razor is
the Levin complexity $Kt$.

\subsection{$S$-Based Predictions}
\label{Spredictions}

If our universe is indeed sampled from the speed prior $S$,
then we might well be able to discover the algorithm for the
apparent noise on top of the seemingly simple physical laws  
--- compare Example \ref{universe}.
It may not be trivial, as trivial pseudorandom generators (PRGs) may not
be quite sufficient for evolution of observers such as ourselves, given
the other laws of physics. But it should be much less time-consuming
than, say, an algorithm computing the $z$ of Theorem \ref{z}
which are effectively indistinguishable from true, incompressible noise.

Based on prior $S$, we predict: anything that appears random or noisy
in our own particular world is due to hitherto unknown regularities
that relate seemingly disconnected events to each other via some simple
algorithm that is not only short (the short algorithms are favored by
all describable measures above) but also {\em fast.} This immediately
leads to more specific predictions.

\subsubsection{Beta Decay} 
When exactly will a particular neutron decay into
a proton, an electron and an antineutrino?  Is the moment of its death
correlated with other events in our universe? Conventional wisdom rejects
this idea and suggests that beta decay is a source of true randomness.
According to $S$, however, this cannot be the case.  Never-ending
true randomness is neither formally describable 
(Def. \ref{describability}) nor S-describable (Def. \ref{quickly}); 
its computation would not be possible using countable 
computational steps.

This encourages a re-examination of beta decay or other types of particle
decay: given $S$, a very simple {\em and} fast but maybe not quite trivial 
PRG should be responsible for the decay pattern of possibly 
widely separated neutrons. (If the PRG were {\em too} trivial and
too obvious then maybe the resulting universe would be too simple to
permit evolution of our type of consciousness, thus being ruled
out by the weak anthropic principle.)  Perhaps the main reason for the
current absence of empirical evidence in this vein is that nobody has
systematically looked for it yet.

\subsubsection{Many World Splits} 
Everett's many worlds hypothesis
\cite{Everett:57} essentially states: whenever our universe's
quantum mechanics based on Schr\"{o}dinger's equation allows for
alternative ``collapses of the wave function,'' {\em all} are made
and the world splits into separate universes.  The previous paper
\cite{Schmidhuber:97brauer} already pointed out that from our algorithmic
point of view there are no real splits --- there are just a bunch of
different algorithms which yield identical results for some time, until
they start computing different outputs corresponding to different possible
observations in different universes.  According to $P^G,P^E,\mu^E,\mu^M$,
$S$, however, most of these alternative continuations are much less
likely than others.

In particular, the outcomes of experiments involving entangled states,
such as the observations of spins of initially close but soon distant
particles with correlated spins, are currently widely assumed to be random. 
Given $S$, however, whenever there are several possible continuations of our
universe corresponding to different wave function collapses, 
and all are compatible with whatever it is we call our
consciousness, we are more likely to end up in one computable 
by a short {\em and} fast algorithm.  A re-examination of split
experiment data might reveil unexpected, nonobvious, nonlocal
algorithmic regularity due to a PRG.

This prediction runs against current mainstream trends in physics,
with the possible exception of hidden variable theory,
e.g., \cite{Bell:66,Bohm:93,Hooft:99}.

\subsubsection{Expected Duration of the Universe} 
Given $S$, the probability that the
history of the universe so far will reach into the next phase of {\bf
FAST} is at most $\frac{1}{2}$  --- compare Example \ref{duration}.
Does that mean there is a 50 \% chance that our universe will get
at least twice as old as it is now? Not necessarily, if the computation
of its state at the $n$-th time step (local time) 
requires more than $O(n)$ time.

As long as there is no compelling contrarian evidence, however, a
reasonable guess would be that our universe is indeed among the fastest
ones with $O(1)$ output bits per constant time interval consumed by
algorithm {\bf FAST}.  It may even be ``locally'' computable through
simple simulated processors,  each interacting with only few neighbouring
processors, assuming that the pseudorandom aspects of our universe
do not require any more global communication between spatio-temporally
separated parts than the well-known physical laws.  Note that the fastest
universe evolutions include those representable as sequences of substrings
of constant length $l$, where each substring stands for the universe's
discretized state at a certain discrete time step and is computable from
the previous substring in $O(l)$ time (compare Example \ref{universe}).
However, the fastest universes also include those whose representations of
successive discrete time steps do grow over time and where more and more
time is spent on their computation.  The expansion of certain computable
universes actually requires this.

In any case, the probability that ours will last $2^n$ times longer
than it has lasted so far is at most $2^{-n}$ (except, of course, when
its early states are for some reason much harder to compute than later
ones {\em and} we are still in an early state).  This prediction also
differs from those of current mainstream physics (compare \cite{Gott:93}
though), but obviously is not verifiable.

\subsection{Short Algorithm Detectable?}

Simple PRG subroutines of the universe may not necessarily be easy to
find.  For instance, the second billion bits of $\pi$'s dyadic expansion
``look'' highly random although they are not, because they are computable
by a very short algorithm.  Another problem with existing data may be
its potential incompleteness. To exemplify this: it is easy to see the
pattern in an observed sequence $1, 2, 3, \ldots, 100$. But if many
values are missing, resulting in an observed subsequence of, say, $7,
19, 54, 57$, the pattern will be less obvious.

A systematic enumeration and execution of all candidate algorithms
in the time-optimal style of Levin search \cite{Levin:73} should find
one consistent with the data essentially as quickly as possible.
Still, currently we do not have an {\em a priori} upper bound on the search
time. This points to a problem of falsifiability.

Another caveat is that the algorithm computing our universe may somehow
be wired up to defend itself against the discovery of its simple PRG.
According to Heisenberg we cannot observe the precise, current state of
a single electron, let alone our universe, because our actions seem to
influence our measurements in a fundamentally unpredictable way.  This
does not rule out a predictable underlying computational process whose
deterministic results we just cannot access \cite{Schmidhuber:97brauer}
--- compare hidden variable theory \cite{Bell:66,Bohm:93,Hooft:99}.
More research, however, is necessary to determine to what extent such
fundamental undetectability is possible in principle from a computational
perspective (compare \cite{Svozil:94,Roessler:98}).

For now there is no reason why believers in $S$ should let
themselves get discouraged too quickly from searching for simple
algorithmic regularity in apparently noisy physical events such
as beta decay and ``many world splits'' in the spirit of Everett
\cite{Everett:57}.  The potential rewards of such a revolutionary
discovery would merit significant experimental and analytic efforts.

\subsection{Relation to Previous Work on All Possible Universes}

A previous paper on computable universes \cite{Schmidhuber:97brauer}
already pointed out that computing all universes with all possible
types of physical laws tends to be much cheaper in terms of information
requirements than computing just one particular, arbitrarily chosen
one, because there is an extremely short algorithm that systematically
enumerates and runs {\em all} computable universes, while most individual universes
have very long shortest descriptions.  The subset embodied by the
many worlds of Everett III's ``many worlds hypothesis'' \cite{Everett:57}
was considered a by-product of this more general set-up.

The previous paper apparently also was the first to apply the theory
of inductive inference to entire universe histories \cite[Section:
{\em Are we Run by a Short Algorithm?}]{Schmidhuber:97brauer}, using
the Solomonoff-Levin distribution to predict that simple universes are
more likely; that is, the most probable universe is the simplest one
compatible with our existence, where simplicity is defined in terms
of traditional Kolmogorov complexity --- compare {\em everything mailing
list archive:} {\em http://www.escribe.com/science/theory/m1284.html}
and {\em m1312.html}, as well as recent papers by Standish and Soklakov
\cite{Standish:00,Soklakov:00}, and see Calude and Meyerstein \cite{Calude:99} 
for a somewhat contrarian view.

The current paper introduces
simplicity measures more dominant than the traditional ones
\cite{Kolmogorov:65,Solomonoff:64,Solomonoff:78,Chaitin:69,Zvonkin:70,Levin:73a,Levin:74,Gacs:74,Chaitin:75,Gacs:83,Schnorr:73,Chaitin:87,Gellmann:95,LiVitanyi:97},
and provides a more general, more technical, and more detailed account,
incorporating several novel theoretical results based on generalizations
of Kolmogorov complexity and algorithmic probability.  In particular, it
stretches the notions of computability and constructivism to the limits,
by considering not only MTM-based traditional computability but also less
restrictive GTM-based and EOM-based describability, and proves several
relevant ``Occams razor theorems.''  Unlike the previous 
paper \cite{Schmidhuber:97brauer} it also
analyzes fundamental time constraints on the computation of everything,
and derives predictions based on these restrictions.

Rather than pursuing the computability-oriented path layed out in
\cite{Schmidhuber:97brauer}, Tegmark recently suggested what at first
glance seems to be an alternative ensemble of possible universes based
on a (somewhat vaguely defined) set of ``self-consistent mathematical
structures'' \cite{Tegmark:98}, thus going beyond his earlier, less
general work \cite{Tegmark:96} on physical constants and Everett's many
world variants \cite{Everett:57} of our own particular universe ---
compare also Marchal's and Bostrom's theses \cite{Marchal:98,Bostrom:00}.
It is not quite clear whether Tegmark would like to include universes that
are {\em not} formally describable according to Def. \ref{describability}.
It is well-known, however, that for any set of mathematical axioms
there is a program that lists all provable theorems in order of the
lengths of their shortest proofs encoded as bitstrings.  Since the TM
that computes all bitstrings outputs all these proofs for all possible
sets of axioms, Tegmark's view \cite{Tegmark:98} seems in a certain sense
encompassed by the algorithmic approach \cite{Schmidhuber:97brauer}. On
the other hand, there are many formal axiomatic systems powerful enough
to encode all computations of all possible TMs, e.g., number theory.
In this sense the algorithmic approach is encompassed by number theory.

The algorithmic approach, however, offers several conceptual
advantages: (1) It provides the appropriate framework for issues
of information-theoretic complexity traditionally ignored in pure
mathematics, and imposes natural complexity-based orderings on the
possible universes and subsets thereof.  (2) It taps into a rich
source of theoretical insights on computable probability distributions
relevant for establishing priors on possible universes. Such priors
are needed for making probabilistic predictions concerning our own
particular universe. Although Tegmark suggests that {\em ``... all
mathematical structures are a priori given equal statistical weight''}
\cite{Tegmark:98}(p. 27), there is no way of assigning equal nonvanishing
probability to all (infinitely many) mathematical structures. Hence
we really need something like the complexity-based weightings discussed in
\cite{Schmidhuber:97brauer} and especially the paper at hand.  (3) The
algorithmic approach is the obvious framework for questions of temporal
complexity such as those discussed in this paper, e.g.,
``what is the most efficient way of simulating all universes?''

\section{Concluding Remarks}
\label{conclusion}

There is an entire spectrum of ways of ordering the describable
things, spanned by two extreme ways of doing it.  Sections
\ref{preliminaries}-\ref{coding} analyzed one of the extremes,
based on minimal constructive description size on generalized Turing Machines more
expressive than those considered in previous work on Kolmogorov
complexity and algorithmic probability and inductive inference.
Section \ref{temporalcomp} discussed the other extreme based on the
fastest way of computing all computable things.

Between the two extremes we find methods for ordering describable
things by (a) their minimal nonhalting {\em enumerable} descriptions
(also discussed in Sections \ref{preliminaries}-\ref{coding}),
(b) their minimal {\em halting} or monotonic descriptions (this is the traditional 
theory of Kolmogorov complexity or algorithmic information),
and (c) the polynomial time complexity-oriented criteria
being subject of most work in theoretical computer science. 
Theorems in Sections \ref{preliminaries}-\ref{temporalcomp} 
reveil some of the structure of the computable and enumerable and
constructively describable things.

Both extremes of the spectrum as well as some of the intermediate
points yield natural prior distributions on describable objects.
The approximable and cumulatively enumerable description size-based
priors (Sections \ref{measures}-\ref{coding}) suggest algorithmic
theories of everything (TOEs) partially justifying Occam's razor in a
way more general than previous approaches: given several explanations
of your universe, those requiring few bits of information are much more
probable than those requiring many bits (Section \ref{consequences}).
However, there may not be an effective procedure for discovering a
compact and complete explanation even if there is one.

The resource-optimal, less dominant, yet arguably more plausible
extreme (Section \ref{temporalcomp}) leads to an algorithmic TOE without
excessive temporal complexity: no calculation of any universe computable
in countable time needs to suffer from an essential slow-down due to
simultaneous computation of all the others.  Based on the rather weak
assumption that the world's creator is constrained by certain limits of
computability, and considering that all of us may be just accidental
by-products of His optimally efficient search for a solution to some
computational problem, the  resulting ``speed prior'' predicts that a
fast and short algorithm is responsible not only for the apparently
simple laws of physics but even for what most physicists currently
classify as noise or randomness (Section \ref{consequences}).  It may
be not all that hard to find; we should search for it.

Much of this paper highlights differences between countable
and uncountable sets.  It is argued (Sections \ref{temporalcomp},
\ref{consequences}) that things such as {\em un}countable time and space
and {\em in}computable probabilities actually should not play a role in
explaining the world, for lack of evidence that they are really necessary.
Some may feel tempted to counter this line of reasoning by pointing out
that for centuries physicists have calculated with continua of real
numbers, most of them incomputable.  Even quantum physicists who are
ready to give up the assumption of a continuous universe usually do
take for granted the existence of continuous probability distributions
on their discrete universes, and Stephen Hawking explicitly said: {\em
``Although there have been suggestions that space-time may have a
discrete structure I see no reason to abandon the continuum theories
that have been so successful.''} Note, however, that all physicists in
fact have only manipulated discrete symbols, thus generating finite,
describable proofs of their results derived from enumerable axioms.
That real numbers really {\em exist} in a way transcending the finite
symbol strings used by everybody may be a figment of imagination ---
compare Brouwer's constructive mathematics \cite{Brouwer:07,Beeson:85}
and the L\"{o}wenheim-Skolem Theorem \cite{Loewenheim:15,Skolem:19}
which implies that any first order theory with an uncountable model
such as the real numbers also has a countable model.  As Kronecker put
it: {\em ``Die ganze Zahl schuf der liebe Gott, alles \"{U}brige ist
Menschenwerk''} (``God created the integers, all else is the work of
man'' \cite{Cajori:19}).  Kronecker greeted with scepticism Cantor's
celebrated insight \cite{Cantor:1874} about real numbers,
mathematical objects Kronecker believed did not even exist.

A good reason to study algorithmic, noncontinuous, discrete TOEs is
that they are the simplest ones compatible with everything we know,
in the sense that universes that cannot even be described formally are
obviously less simple than others. In particular, the speed prior-based
algorithmic TOE (Sections \ref{temporalcomp}, \ref{consequences}) neither
requires an uncountable ensemble of universes (not even describable in the
sense of Def. \ref{quickly}), nor infinitely many bits to specify
nondescribable real-valued probabilities or nondescribable infinite random
sequences. One may believe in the validity of algorithmic TOEs until
(a) there is evidence against them, e.g., someone shows that our own
universe is not formally describable and would not be possible without,
say, existence of incomputable numbers, or (b) someone comes up with an
even simpler explanation of everything.  But what could that possibly be?

Philosophers tend to create theories inspired by recent scientific
developments.  For instance, Heisenberg's uncertainty principle
and G\"{o}del's incompleteness theorem greatly influenced modern
philosophy.  Are algorithmic TOEs and the ``Great Programmer
Religion'' \cite{Schmidhuber:97brauer} just another reaction to recent
developments, some in hindsight obvious by-product of the advent of
good virtual reality?  Will they soon become obsolete, as so many
previous philosophies? We find it hard to imagine so, even without a
boost to be expected for algorithmic TOEs in case someone should indeed
discover a simple subroutine responsible for certain physical events
hitherto believed to be irregular. After all, algorithmic theories of
the describable do encompass everything we will ever be able to talk
and write about. Other things are simply beyond description.

\subsection*{Acknowledgments}

At the age of 17 my brother Christof declared that the universe is
a mathematical structure inhabited by observers who are mathematical
substructures (private communication, Munich, 1981).  As he went on to
become a theoretical physicist, discussions with him about the relation
between superstrings and bitstrings became a source of inspiration
for writing both the earlier paper \cite{Schmidhuber:97brauer} and the
present one, both based on computational complexity theory, which seems
to provide the natural setting for his more physics-oriented ideas
(private communication, Munich 1981-86; Pasadena 1987-93; Princeton
1994-96; Berne/Geneva 1997--; compare his notion of {\em ``mathscape''}
\cite{Christof:00}).  Furthermore, Christof's 1997 remarks on similarities
and differences between Feynman path integrals and ``the sum of all
computable universes'' and his resulting dissatisfaction with the lack of
a discussion of temporal aspects in \cite{Schmidhuber:97brauer} triggered
Section \ref{temporalcomp} on temporal complexity.

I am grateful to Ray Solomonoff for his helpful comments on earlier
work \cite{Schmidhuber:97nn}  making use of the probabilistic algorithm
of Section \ref{temporalcomp}, and to Paul Vit\'{a}nyi for useful
information relevant to the proof of Theorem \ref{nouniversal}.  I would
also like to express my thanks to numerous posters and authors (e.g.,
\cite{Marchal:98,Tegmark:98,Moravec:99,Bostrom:00,Standish:00,Donald:90,Higgo:99,Mallah:00,Ruhl:00})
of the {\em everything mailing list} created by Wei Dai \cite{Dai:98}
{\em (everything-list@eskimo.com)}.  Some of the text above actually
derives from my replies to certain postings (see archive at {\em
http://www.escribe.com/science/theory/}).  
Finally I am indebted 
to Marcus Hutter and Sepp Hochreiter for independently checking the theorems,
to Leonora Bianchi, 
Wei Dai,
Doug Eck, 
Felix Gers, 
Ivo Kwee, 
Carlo Lepori, 
Leonid Levin,
Monaldo Mastrolilli,
Andrea Rizzoli, 
Nicol N. Schraudolph, 
and Marco Zaffalon, 
for comments on (parts of) earlier drafts 
or of Version 1.0 \cite{Schmidhuber:00version1},
to Wilfried Brauer and Karl Svozil for relevant pointers and references,
and especially to Marcus Hutter for the proof of Theorem \ref{nouniversal}.

%\bibliography{bib}

\begin{thebibliography}{100}

\bibitem{Adleman:79}
L.~Adleman.
\newblock Time, space, and randomness.
\newblock Technical Report MIT/LCS/79/TM-131, Laboratory for Computer Science,
  MIT, 1979.

\bibitem{Allender:92}
A.~Allender.
\newblock Application of time-bounded {Kolmogorov} complexity in complexity
  theory.
\newblock In O.~Watanabe, editor, {\em {Kolmogorov} complexity and
  computational complexity}, pages 6--22. EATCS Monographs on Theoretical
  Computer Science, Springer, 1992.

\bibitem{Allender:89}
E.~Allender.
\newblock Some consequences of the existence of pseudorandom generators.
\newblock {\em Journal of Computer and System Science}, 39:101--124, 1989.

\bibitem{BarrowTipler:86}
J.~D. Barrow and F.~J. Tipler.
\newblock {\em The Anthropic Cosmological Principle}.
\newblock Clarendon Press, Oxford, 1986.

\bibitem{Barzdin:88}
Y.~M. Barzdin.
\newblock Algorithmic information theory.
\newblock In D.~Reidel, editor, {\em Encyclopaedia of Mathematics}, volume~1,
  pages 140--142. Kluwer Academic Publishers, 1988.

\bibitem{Beeson:85}
M.~Beeson.
\newblock {\em Foundations of Constructive Mathematics}.
\newblock Springer-Verlag, Heidelberg, 1985.

\bibitem{Bell:66}
J.~S. Bell.
\newblock On the problem of hidden variables in quantum mechanics.
\newblock {\em Rev. Mod. Phys.}, 38:447--452, 1966.

\bibitem{Bennett:88}
C.~H. Bennett.
\newblock Logical depth and physical complexity.
\newblock In {\em The Universal {Turing} Machine: A Half Century Survey},
  volume~1, pages 227--258. Oxford University Press, Oxford and Kammerer \&
  Unverzagt, Hamburg, 1988.

\bibitem{Bennett:00}
C.~H. Bennett and D.~P. DiVicenzo.
\newblock Quantum information and computation.
\newblock {\em Nature}, 404(6775):256--259, 2000.

\bibitem{Blum:89}
L.~Blum, M.~Shub, and S.~Smale.
\newblock On a theory of computation and complexity over the real numbers: {NP}
  completeness, recursive functions, and universal machines.
\newblock {\em Bulletin AMS}, 21, 1989.

\bibitem{Blumer:87}
A.~Blumer, A.~Ehrenfeucht, D.~Haussler, and M.~K. Warmuth.
\newblock Occam's razor.
\newblock {\em Information Processing Letters}, 24:377--380, 1987.

\bibitem{Bohm:93}
D.~Bohm and B.~J. Hiley.
\newblock {\em The Undivided Universe}.
\newblock Routledge, New York, N.Y., 1993.

\bibitem{Boskovich:1755}
R.~J. Boskovich.
\newblock {\em De spacio et tempore, ut a nobis cognoscuntur}.
\newblock Vienna, 1755.
\newblock English translation in \cite{Boskovich:1966}.

\bibitem{Boskovich:1966}
R.~J. Boskovich.
\newblock De spacio et tempore, ut a nobis cognoscuntur.
\newblock In J.~M. Child, editor, {\em A Theory of Natural Philosophy}, pages
  203--205. Open Court (1922) and MIT Press, Cambridge, MA, 1966.

\bibitem{Bostrom:00}
N.~Bostrom.
\newblock {Observational selection effects and probability. Dissertation, Dept.
  of Philosophy, Logic and Scientific Method, London School of Economics},
  2000.

\bibitem{Bremermann:82}
H.~J. Bremermann.
\newblock Minimum energy requirements of information transfer and computing.
\newblock {\em International Journal of Theoretical Physics}, 21:203--217,
  1982.

\bibitem{Brouwer:07}
L.~E.~J. Brouwer.
\newblock {Over de Grondslagen der Wiskunde. Dissertation, Doctoral Thesis,
  University of Amsterdam}, 1907.

\bibitem{Burgin:83}
M.~S. Burgin.
\newblock Inductive {Turing} machines.
\newblock {\em Notices of the Academy of Sciences of the USSR (translated from
  {Russian})}, 270(6):1289--1293, 1991.

\bibitem{Burgin:91}
M.~S. Burgin and Y.~M. Borodyanskii.
\newblock Infinite processes and super-recursive algorithms.
\newblock {\em Notices of the Academy of Sciences of the USSR (translated from
  {Russian})}, 321(5):800--803, 1991.

\bibitem{Cajori:19}
F.~Cajori.
\newblock {\em History of mathematics (2nd edition)}.
\newblock Macmillan, New York, 1919.

\bibitem{Calude:00}
C.~S. Calude.
\newblock {Chaitin $\Omega$ numbers, Solovay machines and G\"{o}del
  incompleteness}.
\newblock {\em Theoretical Computer Science}, 2000.
\newblock In press.

\bibitem{Calude:99}
C.~S. Calude and F.~W. Meyerstein.
\newblock Is the universe lawful?
\newblock {\em Chaos, Solitons \& Fractals}, 10(6):1075--1084, 1999.

\bibitem{Cantor:1874}
G.~Cantor.
\newblock {\"{U}ber eine Eigenschaft des Inbegriffes aller reellen
  algebraischen Zahlen}.
\newblock {\em Crelle's Journal f\"{u}r Mathematik}, 77:258--263, 1874.

\bibitem{Carter:74}
B.~Carter.
\newblock Large number coincidences and the anthropic principle in cosmology.
\newblock In M.~S. Longair, editor, {\em Proceedings of the IAU Symposium 63},
  pages 291--298. Reidel, Dordrecht, 1974.

\bibitem{Hotz:99}
T.~Chadzelek and G.~Hotz.
\newblock Analytic machines.
\newblock {\em Theoretical Computer Science}, 219:151--167, 1999.

\bibitem{Chaitin:69}
G.J. Chaitin.
\newblock On the length of programs for computing finite binary sequences:
  statistical considerations.
\newblock {\em Journal of the ACM}, 16:145--159, 1969.
\newblock Submitted 1965.

\bibitem{Chaitin:75}
G.J. Chaitin.
\newblock A theory of program size formally identical to information theory.
\newblock {\em Journal of the ACM}, 22:329--340, 1975.

\bibitem{Chaitin:87}
G.J. Chaitin.
\newblock {\em Algorithmic Information Theory}.
\newblock Cambridge University Press, Cambridge, 1987.

\bibitem{Cover:89}
T.~M. Cover, P.~G\'{a}cs, and R.~M. Gray.
\newblock Kolmogorov's contributions to information theory and algorithmic
  complexity.
\newblock {\em Annals of Probability Theory}, 17:840--865, 1989.

\bibitem{Dai:98}
W.~Dai.
\newblock Everything mailing list archive at
  http://www.escribe.com/science/theory/, 1998.

\bibitem{Deutsch:97}
D.~Deutsch.
\newblock {\em The Fabric of Reality}.
\newblock Allen Lane, New York, NY, 1997.

\bibitem{Donald:90}
M.~J. Donald.
\newblock Quantum theory and the brain.
\newblock {\em Proceedings of the Royal Society (London) Series A}, 427:43--93,
  1990.

\bibitem{Everett:57}
H.~{Everett III}.
\newblock {`Relative State'} formulation of quantum mechanics.
\newblock {\em Reviews of Modern Physics}, 29:454--462, 1957.

\bibitem{Freyvald:74}
R.~V. Freyvald.
\newblock Functions and functionals computable in the limit.
\newblock {\em Transactions of Latvijas Vlasts Univ. Zinatn. Raksti},
  210:6--19, 1977.

\bibitem{Gacs:74}
P.~G\'{a}cs.
\newblock On the symmetry of algorithmic information.
\newblock {\em Soviet Math. Dokl.}, 15:1477--1480, 1974.

\bibitem{Gacs:83}
P.~G\'{a}cs.
\newblock On the relation between descriptional complexity and algorithmic
  probability.
\newblock {\em Theoretical Computer Science}, 22:71--93, 1983.

\bibitem{Gellmann:95}
M.~Gell-Mann.
\newblock Remarks on simplicity and complexity.
\newblock {\em Complexity}, 1(1):16--19, 1995.

\bibitem{Goedel:31}
K.~G\"{o}del.
\newblock \"{U}ber formal unentscheidbare {S\"{a}tze der Principia Mathematica
  und verwandter Systeme I}.
\newblock {\em Monatshefte f\"{u}r Mathematik und Physik}, 38:173--198, 1931.

\bibitem{Gold:65}
E.~M. Gold.
\newblock Limiting recursion.
\newblock {\em Journal of Symbolic Logic}, 30(1):28--46, 1965.

\bibitem{Gott:93}
J.~R. {Gott, III}.
\newblock Implications of the {Copernican} principle for our future prospects.
\newblock {\em Nature}, 363:315--319, 1993.

\bibitem{Greg:57}
A.~Gregorczyk.
\newblock On the definitions of computable real continuous functions.
\newblock {\em Fundamenta Mathematicae}, 44:61--71, 1957.

\bibitem{Hartmanis:83}
J.~Hartmanis.
\newblock Generalized {Kolmogorov} complexity and the structure of feasible
  computations.
\newblock In {\em Proc. 24th IEEE Symposium on Foundations of Computer
  Science}, pages 439--445, 1983.

\bibitem{Higgo:99}
J.~Higgo.
\newblock Physics of enlightenment.
\newblock {\em Middle Way Journal}, February 1999.

\bibitem{Hochreiter:97nc1}
S.~Hochreiter and J.~Schmidhuber.
\newblock Flat minima.
\newblock {\em Neural Computation}, 9(1):1--42, 1997.

\bibitem{Hotz:95}
G.~Hotz, G.~Vierke, and B.~Schieffer.
\newblock Analytic machines.
\newblock Technical Report TR95-025, Electronic Colloquium on Computational
  Complexity, 1995.
\newblock http://www.eccc.uni-trier.de/eccc/.

\bibitem{Huffman:52}
D.~A. Huffman.
\newblock A method for construction of minimum-redundancy codes.
\newblock {\em Proceedings IRE}, 40:1098--1101, 1952.

\bibitem{Hutter:99}
M.~Hutter.
\newblock New error bounds for {Solomonoff} prediction.
\newblock {\em Journal of Computer and System Science, in press}, 2000.
\newblock http://xxx.lanl.gov/abs/cs.AI/9912008.

\bibitem{Hutter:00e}
M.~Hutter.
\newblock Optimality of universal prediction for general loss and alphabet.
\newblock Technical report, Istituto Dalle Molle di Studi sull'Intelligenza
  Artificiale, Manno (Lugano), CH, December 2000.
\newblock In progress.

\bibitem{Hutter:00f}
M.~Hutter.
\newblock A theory of universal artificial intelligence based on algorithmic
  complexity.
\newblock Technical Report IDSIA-14-00 (cs.AI/0004001), IDSIA, Manno (Lugano),
  CH, 2000.
\newblock http://xxx.lanl.gov/abs/cs.AI/0004001.

\bibitem{Kolmogorov:65}
A.N. Kolmogorov.
\newblock Three approaches to the quantitative definition of information.
\newblock {\em Problems of Information Transmission}, 1:1--11, 1965.

\bibitem{Kraft:49}
L.~G. Kraft.
\newblock {A device for quantizing, grouping, and coding amplitude modulated
  pulses. M.Sc. Thesis, Dept. of Electrical Engineering, MIT, Cambridge,
  Mass.}, 1949.

\bibitem{Levin:73a}
L.~A. Levin.
\newblock On the notion of a random sequence.
\newblock {\em Soviet Math. Dokl.}, 14(5):1413--1416, 1973.

\bibitem{Levin:73}
L.~A. Levin.
\newblock Universal sequential search problems.
\newblock {\em Problems of Information Transmission}, 9(3):265--266, 1973.

\bibitem{Levin:74}
L.~A. Levin.
\newblock Laws of information (nongrowth) and aspects of the foundation of
  probability theory.
\newblock {\em Problems of Information Transmission}, 10(3):206--210, 1974.

\bibitem{Levin:84}
L.~A. Levin.
\newblock Randomness conservation inequalities: Information and independence in
  mathematical theories.
\newblock {\em Information and Control}, 61:15--37, 1984.

\bibitem{LiVitanyi:97}
M.~Li and P.~M.~B. Vit\'{a}nyi.
\newblock {\em An Introduction to {Kolmogorov} Complexity and its Applications
  (2nd edition)}.
\newblock Springer, 1997.

\bibitem{Lloyd:00}
S.~Lloyd.
\newblock Ultimate physical limits to computation.
\newblock {\em Nature}, 406:1047--1054, 2000.

\bibitem{Loewenheim:15}
L.~L\"{o}wenheim.
\newblock {\"{U}ber M\"{o}glichkeiten im Relativkalk\"{u}l}.
\newblock {\em Mathematische Annalen}, 76:447--470, 1915.

\bibitem{Mallah:00}
J.~Mallah.
\newblock The computationalist wavefunction interpretation agenda {(CWIA)}.
  {Continually} modified draft, Dec 2000.
\newblock http://hammer.prohosting.com/\~{ }mathmind/cwia.htm {\em
  (nonpermanent contents)}.

\bibitem{Marchal:98}
B.~Marchal.
\newblock {\em Calculabilit\'{e}, Physique et Cognition}.
\newblock PhD thesis, L'Universit\'{e} des Sciences et Technologies De Lilles,
  1998.

\bibitem{Martin-Loef:66}
P.~Martin-L\"{o}f.
\newblock The definition of random sequences.
\newblock {\em Information and Control}, 9:602--619, 1966.

\bibitem{Moravec:99}
H.~Moravec.
\newblock {\em Robot}.
\newblock Wiley Interscience, 1999.

\bibitem{Mostowski:57}
A.~Mostowski.
\newblock On computable sequences.
\newblock {\em Fundamenta Mathematicae}, 44:37--51, 1957.

\bibitem{Penrose:89}
R.~Penrose.
\newblock {\em The Emperor's New Mind.}
\newblock Oxford University Press, 1989.

\bibitem{Putnam:65}
H.~Putnam.
\newblock Trial and error predicates and the solution to a problem of
  {Mostowski}.
\newblock {\em Journal of Symbolic Logic}, 30(1):49--57, 1965.

\bibitem{Rissanen:86}
J.~Rissanen.
\newblock Stochastic complexity and modeling.
\newblock {\em The Annals of Statistics}, 14(3):1080--1100, 1986.

\bibitem{Rogers:67}
H.~{Rogers, Jr.}
\newblock {\em Theory of Recursive Functions and Effective Computability}.
\newblock McGraw-Hill, New York, 1967.

\bibitem{Roessler:98}
Otto~E. R{\"{o}}ssler.
\newblock {\em Endophysics. {T}he World as an Interface}.
\newblock World Scientific, Singapore, 1998.
\newblock With a foreword by Peter Weibel.

\bibitem{Ruhl:00}
H.~Ruhl.
\newblock The use of complexity to solve dilemmas in physics. {Continually}
  modified draft, Dec 2000.
\newblock http://www.connix.com/\~{ }hjr/model01.html {\em (nonpermanent
  contents)}.

\bibitem{Christof:00}
C.~Schmidhuber.
\newblock Strings from logic.
\newblock Technical Report CERN-TH/2000-316, CERN, Theory Division, 2000.
\newblock http://xxx.lanl.gov/abs/hep-th/0011065.

\bibitem{Schmidhuber:95kol}
J.~Schmidhuber.
\newblock Discovering solutions with low {Kolmogorov} complexity and high
  generalization capability.
\newblock In A.~Prieditis and S.~Russell, editors, {\em Machine Learning:
  Proceedings of the Twelfth International Conference}, pages 488--496. Morgan
  Kaufmann Publishers, San Francisco, CA, 1995.

\bibitem{Schmidhuber:97brauer}
J.~Schmidhuber.
\newblock A computer scientist's view of life, the universe, and everything.
\newblock In C.~Freksa, M.~Jantzen, and R.~Valk, editors, {\em Foundations of
  Computer Science: Potential - Theory - Cognition}, volume 1337, pages
  201--208. Lecture Notes in Computer Science, Springer, Berlin, 1997.
\newblock Submitted 1996.

\bibitem{Schmidhuber:97nn}
J.~Schmidhuber.
\newblock Discovering neural nets with low {Kolmogorov} complexity and high
  generalization capability.
\newblock {\em Neural Networks}, 10(5):857--873, 1997.

\bibitem{Schmidhuber:97art}
J.~Schmidhuber.
\newblock Low-complexity art.
\newblock {\em Leonardo, Journal of the International Society for the Arts,
  Sciences, and Technology}, 30(2):97--103, 1997.

\bibitem{Schmidhuber:00version1}
J.~Schmidhuber.
\newblock Algorithmic theories of everything.
\newblock Technical Report IDSIA-20-00, Version 1.0, IDSIA, Manno (Lugano),
  Switzerland, November 2000.
\newblock http://arXiv.org/abs/quant-ph/0011122.

\bibitem{Schmidhuber:97bias}
J.~Schmidhuber, J.~Zhao, and M.~Wiering.
\newblock Shifting inductive bias with success-story algorithm, adaptive
  {Levin} search, and incremental self-improvement.
\newblock {\em Machine Learning}, 28:105--130, 1997.

\bibitem{Schnorr:73}
C.~P. Schnorr.
\newblock Process complexity and effective random tests.
\newblock {\em Journal of Computer Systems Science}, 7:376--388, 1973.

\bibitem{Shannon:48}
C.~E. Shannon.
\newblock A mathematical theory of communication (parts {I} and {II}).
\newblock {\em Bell System Technical Journal}, XXVII:379--423, 1948.

\bibitem{Skolem:19}
T.~Skolem.
\newblock {Logisch-kombinatorische Untersuchungen \"{u}ber Erf\"{u}llbarkeit
  oder Beweisbarkeit mathematischer S\"{a}tze nebst einem Theorem \"{u}ber
  dichte Mengen}.
\newblock {\em Skrifter utgit av Videnskapsselskapet in Kristiania, I,
  Mat.-Nat. Kl.}, N4:1--36, 1919.

\bibitem{Slaman:99}
T.~Slaman.
\newblock Randomness and recursive enumerability.
\newblock Technical report, Univ. of California, Berkeley, 1999.
\newblock Preprint, http://www.math.berkeley.edu/\~{ }slaman.

\bibitem{Soklakov:00}
A.~N. Soklakov.
\newblock Occam's razor as a formal basis for a physical theory.
\newblock Technical Report math-ph/0009007, Univ. London, Dept. Math., Royal
  Holloway, Egham, Surrey TW20 OEX, September 2000.
\newblock http://arXiv.org/abs/math-ph/0009007.

\bibitem{Solomonoff:64}
R.J. Solomonoff.
\newblock A formal theory of inductive inference. {Part I}.
\newblock {\em Information and Control}, 7:1--22, 1964.

\bibitem{Solomonoff:78}
R.J. Solomonoff.
\newblock Complexity-based induction systems.
\newblock {\em IEEE Transactions on Information Theory}, IT-24(5):422--432,
  1978.

\bibitem{Solovay:75}
R.~M. Solovay.
\newblock Lecture notes on algorithmic complexity, {UCLA}, unpublished, 1975.

\bibitem{Solovay:00}
R.~M. Solovay.
\newblock A version of {$\Omega$} for which {ZFC} can not predict a single bit.
\newblock In C.~S. Calude and G.~P\u{a}un, editors, {\em Finite Versus
  Infinite. Contributions to an Eternal Dilemma}, pages 323--334. Springer,
  London, 2000.

\bibitem{Standish:00}
R.~Standish.
\newblock Why {Occam's} razor?
\newblock Technical report, High Performance Computing Support Unit, Univ. New
  South Wales, Sydney, 2052, Australia, July 2000.

\bibitem{Svozil:94}
Karl Svozil.
\newblock Extrinsic-intrinsic concept and complementarity.
\newblock In H.~Atmanspacker and G.~J. Dalenoort, editors, {\em Inside versus
  Outside}, pages 273--288. Springer-Verlag, Heidelberg, 1994.

\bibitem{Tegmark:96}
M.~Tegmark.
\newblock Does the universe in fact contain almost no information?
\newblock {\em Foundations of Physics Letters}, 9(1):25--42, 1996.

\bibitem{Tegmark:98}
M.~Tegmark.
\newblock Is ``the theory of everything'' merely the ultimate ensemble theory?
\newblock {\em Annals of Physics}, 270:1--51, 1998.
\newblock Submitted 1996.

\bibitem{Hooft:99}
G.~t'Hooft.
\newblock Quantum gravity as a dissipative deterministic system.
\newblock Technical Report SPIN-1999/07/gr-gc/9903084,
  http://xxx.lanl.gov/abs/gr-qc/9903084, Institute for Theoretical Physics,
  Univ. of Utrecht, and Spinoza Institute, Netherlands, 1999.
\newblock Also published in {\em Classical and Quantum Gravity 16}, 3263.

\bibitem{Toffoli:79}
T.~Toffoli.
\newblock The role of the observer in uniform systems.
\newblock In G.~Klir, editor, {\em Applied General Systems Research}. Plenum
  Press, New York, London, 1978.

\bibitem{Turing:36}
A.~M. Turing.
\newblock On computable numbers, with an application to the
  {Entscheidungsproblem}.
\newblock {\em Proceedings of the London Mathematical Society, Series 2},
  41:230--267, 1936.

\bibitem{Uspensky:92}
V.~A. Uspensky.
\newblock Complexity and entropy: an introduction to the theory of {Kolmogorov}
  complexity.
\newblock In O.~Watanabe, editor, {\em {Kolmogorov} complexity and
  computational complexity}, pages 85--102. EATCS Monographs on Theoretical
  Computer Science, Springer, 1992.

\bibitem{Vyugin:98}
V.~V. V'yugin.
\newblock Non-stochastic infinite and finite sequences.
\newblock {\em Theoretical Computer Science}, 207(2):363--382, 1998.

\bibitem{Wallace:68}
C.~S. Wallace and D.~M. Boulton.
\newblock An information theoretic measure for classification.
\newblock {\em Computer Journal}, 11(2):185--194, 1968.

\bibitem{Watanabe:92}
O.~Watanabe.
\newblock {\em {Kolmogorov} complexity and computational complexity}.
\newblock EATCS Monographs on Theoretical Computer Science, Springer, 1992.

\bibitem{Wiering:96levin}
M.A. Wiering and J.~Schmidhuber.
\newblock Solving {POMDPs} with {L}evin search and {EIRA}.
\newblock In L.~Saitta, editor, {\em Machine Learning: Proceedings of the
  Thirteenth International Conference}, pages 534--542. Morgan Kaufmann
  Publishers, San Francisco, CA, 1996.

\bibitem{Zurek:89b}
W.~H. Zurek.
\newblock Algorithmic randomness and physical entropy {I}.
\newblock {\em Phys. Rev.}, A40:4731--4751, 1989.

\bibitem{Zurek:91}
W.~H. Zurek.
\newblock Decoherence and the transition from quantum to classical.
\newblock {\em Physics Today}, 44(10):36--44, 1991.

\bibitem{Zvonkin:70}
A.~K. Zvonkin and L.~A. Levin.
\newblock The complexity of finite objects and the algorithmic concepts of
  information and randomness.
\newblock {\em Russian Math. Surveys}, 25(6):83--124, 1970.

\end{thebibliography}
\bibliographystyle{plain}

\end{document}